\documentclass[11pt,numbers,sort&compress]{article}

\usepackage{latexsym,amsmath,amssymb,theorem,epsfig} 
\usepackage{natbib}
\usepackage{graphicx}

\DeclareFontFamily{OT1}{rsfs10}{}
\DeclareFontShape{OT1}{rsfs10}{m}{n}{ <-> rsfs10 }{}
\DeclareMathAlphabet{\mathscript}{OT1}{rsfs10}{m}{n}

\newcommand{\be}{\begin{equation}}
\newcommand{\ee}{\end{equation}}
\newcommand{\nn}{\nonumber}
\newcommand{\bea}{\begin{eqnarray}}
\newcommand{\eea}{\end{eqnarray}}
\newcommand{\ba}{\begin{array}}
\newcommand{\ea}{\end{array}}

\newcommand{\tr}{\textrm{tr}}

\newcommand{\pt}{\partial}

\def\a{\alpha}
\def\b{\beta}

\def\d{\delta}

\def\k{\kappa}

\def\w{\wedge}

\def\O{\Omega}

\def\cG{{\cal G}}

\def\cK{{\cal K}}

\begin{document}

\begin{titlepage}
  \begin{flushright}
    hep-th/0701025
  \end{flushright}
  \vspace*{\stretch{1}}
  \begin{center}
     \Large Perturbative Anti-Brane Potentials in Heterotic M-theory
  \end{center}
  \vspace*{\stretch{2}}
  \begin{center}
    \begin{minipage}{\textwidth}
      \begin{center}
        James Gray${}^1$,
        Andr\'e Lukas${}^2$ and
        Burt Ovrut${}^3$
      \end{center}
    \end{minipage}
  \end{center}
  \vspace*{1mm}
  \begin{center}
    \begin{minipage}{\textwidth}
      \begin{center}
        ${}^1$Institut d'Astrophysique de Paris and APC, 
        Universit\'e de Paris 7,\\
        98 bis, Bd.~Arago 75014, Paris, France\\[0.2cm]
        ${}^2$Rudolf Peierls Centre for Theoretical Physics, 
        University of Oxford,\\ 
        1 Keble Road, Oxford OX1 3NP, UK\\[0.2cm]
        ${}^3$Department of Physics, University of Pennsylvania,\\
        Philadelphia, PA 19104--6395, USA
      \end{center}
    \end{minipage}
  \end{center}
  \vspace*{\stretch{1}}
  \begin{abstract}
    \normalsize 
    We derive the perturbative four-dimensional effective theory
    describing heterotic M-theory with branes and anti-branes in the
    bulk space. The back-reaction of both the branes and anti-branes
    is explicitly included. To first order in the heterotic
    $\epsilon_S$ expansion, we find that the forces on branes and
    anti-branes vanish and that the KKLT procedure of simply adding to
    the supersymmetric theory the probe approximation to the energy
    density of the anti-brane reproduces the correct potential. 
    However, there are additional non-supersymmetric
    corrections to the gauge-kinetic functions and matter terms. The
    new correction to the gauge kinetic functions is important in a
    discussion of moduli stabilization. At second order in the
    $\epsilon_S$ expansion, we find that the forces on the branes and
    anti-branes become non-vanishing.  These forces are not precisely
    in the naive form that one may have anticipated and, being second
    order in the small parameter $\epsilon_S$, they are relatively
    weak. This suggests that moduli stabilization in heterotic models
    with anti-branes is achievable.
  \end{abstract}
  \vspace*{\stretch{5}}
  \begin{minipage}{\textwidth}
    \underline{\hspace{5cm}}
    \\
    \footnotesize
    ${}^1$email: gray@iap.fr \\
    ${}^2$email: lukas@physics.ox.ac.uk \\
    ${}^3$email: ovrut@elcapitan.hep.upenn.edu
  \end{minipage}
\end{titlepage}


\section{Introduction}

Recent years have seen considerable progress in various aspects of
string phenomenology, including a better understanding of moduli
stabilization.  By combining a series of different effects,
phenomenologically interesting type II string models with all moduli
stabilized have been found.  The KKLT procedure~\cite{Kachru:2003aw}
first shows that the four- dimensional effective theory associated
with D-branes admits a completely stable supersymmetric AdS
vacuum. Then, an anti D-brane is added to the compactification. This
both breaks supersymmetry and raises the AdS vacuum to a dS one with a
small cosmological constant. Many of the details of this procedure are
still being worked out in the literature; see, for example
~\cite{Grana:2005jc, Conlon:2005ki, Choi:2005ge, Denef:2005mm,
  Balasubramanian:2005zx,Burgess:2006mn}. However, it seems likely
that this is a valid way of obtaining stable de Sitter
non-supersymmetric vacua within the context of IIB string theory.

Heterotic models, on the other hand, offer a number of advantages in
terms of particle physics model building. For example, models with an
underlying ${\rm SO}(10)$ GUT symmetry can be constructed where one
right handed neutrino per family occurs naturally in the $\bf 16 \rm$
multiplet and gauge unification is generic due to the universal gauge
kinetic functions in heterotic theories. Recent progress in the
understanding of non-standard embedding models~\cite{A1,A2,A3,A4} and
the associated mathematics of vector bundles on Calabi-Yau
spaces~\cite{B1,B2,B3,B4} has led to the construction of effective
theories close to the Minimal Supersymmetric Standard Model (MSSM),
see~\cite{Donagi:2004ub, Braun:2005ux, Braun:2005bw, Braun:2005zv,
  Braun:2006ae1, Braun:2006ae}.  This has opened up new avenues for
heterotic phenomenology. For example, one can proceed to look at more
detailed properties of these models such as $\mu$
terms~\cite{Braun:2006ae2}, Yukawa couplings~\cite{Braun:2006ae3}, the
number of moduli~\cite{Braun:2006ae4} and so forth. Many other groups
are also making great strides in model building in heterotic, see for
example
\cite{Andreas:2006dm,Faraggi:2006qa,Lebedev:2006kn,Blumenhagen:2006wj}.

The main motivation of this paper is to combine some of the advantages
of both approaches, type II and heterotic, by realizing key features
of type IIB moduli stabilization within heterotic theories. Related
work on moduli stabilization in heterotic can be found in, for
example,
\cite{LopesCardoso:2002hd,Buchbinder:2003pi,Becker:2004gw,Gurrieri:2004dt,Curio:2005ew,deCarlos:2005kh,Ovrut:20061}.
Specifically, we would like to study heterotic M-theory in the
presence of M five-branes and {\it anti M five-branes}. The starting
point of our analysis is the five-dimensional supersymmetric effective
action of heterotic
M-theory~\cite{Lukas:1998yy,Lukas:1998tt,Brandle:2001ts}, where the M
five-branes appear as three-branes. We will first generalize this
five-dimensional theory to include anti three-branes. Our main goal is
the derivation of the associated perturbative four-dimensional
effective action, including the effects from back-reaction of both the
branes and the anti-branes. As a first step, we find the
five-dimensional non-supersymmetric domain wall in the presence of
anti-branes, a generalization of the BPS domain wall vacuum of the
supersymmetric theory~\cite{Lukas:1998yy,Lukas:1998tt,Brandle:2001ts}.
The four-dimensional effective theory is then obtained as a
dimensional reduction on this domain wall.

Detailed knowledge of this four-dimensional theory is crucial in order
to address a number of important problems in heterotic model building.
Most notably these are moduli stabilization and, in particular, the
stabilization of anti-branes, obtaining a small positive cosmological
constant consistent with observation at the stable minimum and the
nature of supersymmetry breaking due to the anti-branes. There are
also important applications to cosmology analogous to those which have
been seen in the case of moving M5 branes, see for example
\cite{Khoury:2001wf,Copeland:2001zp,Bastero-Gil:2002hs,Copeland:2002fv,deCarlos:2004yv,Becker:2005sg,Copeland:2006hv,Lehners:2006pu,Lehners:2006ir}. These
problems are the main motivation for our work and they will be
explicitly addressed in forthcoming
publications~\cite{otherpapers}. In the present paper, we will
concentrate on the more formal aspect of deriving the relevant
four-dimensional effective theory at the perturbative level. The
relative simplicity of our five-dimensional approach allows us to
explicitly include the effects of the anti-brane back-reaction,
something that is much more difficult to achieve for the complicated
geometries in IIB models with anti-branes. Our results are relevant
for generic string- and M-theory model building with anti-branes.

Let us summarize the main results of this paper. Starting from
five-dimensional heterotic M-theory with three-branes and anti
three-branes, we calculate the associated bosonic four-dimensional
effective theory, including the effects of the back-reaction of the
branes and anti-branes. This is done by dimensional reduction on a
five-dimensional non-supersymmetric domain wall solution which we
explicitly determine. The calculation is performed as a systematic
expansion in powers of $\kappa_{11}^{\frac{2}{3}}$, where
$\kappa_{11}$ is the 11-dimensional Planck constant. To zeroth and
first order we find the effective action is given by the usual
supersymmetric result plus an ``uplifting potential'' from the
anti-branes. This potential is first order in the strong coupling
parameter $\epsilon_S \propto \kappa_{11}^{\frac{2}{3}}$ of heterotic
M-theory and, hence, suppressed. Furthermore, it depends on the
dilaton and the K\"ahler moduli but is, at this order, independent of
the brane position moduli. Hence, to order $\epsilon_S$, the
perturbative force on branes and anti-branes vanishes. This
corroborates and justifies the results of~\cite{Ovrut:20061}, where the
possibility of a meta-stable vacuum with a small positive cosmological
constant was demonstrated within the context of a slope-stable
heterotic standard model. We have also been able to reliably calculate
some contributions to the four-dimensional effective action at order
${\epsilon_S}^{2} \propto \kappa_{11}^{4/3}$. In particular, we have
calculated the complete correction to the brane potential at this
order.  This second order potential describes the expected
``Coulomb-type'' forces between the branes and the anti-branes, but
also contains an additional unexpected interaction between these
objects. This new term can be attributed to the back-reaction of the
anti-branes. It is remarkable that these inter-brane forces are all of
${\cal O}(\epsilon_S^2)$ and are, therefore, strongly suppressed. We
have also calculated the ${\cal O}(\epsilon_S)$ threshold corrections
to the gauge kinetic functions in the presence of anti-branes and find
that they are non-holomorphic due to the supersymmetry breaking by the
anti-branes.  Furthermore, they depend explicitly on the brane and
anti-brane moduli.

Our results have significant implications for heterotic
compactifications with anti-branes, in particular for the problem of
moduli stabilization in such models. The relative weakness of the
perturbative forces between the branes means that it is possible that
branes and anti-branes can be stabilized by balancing the anti-brane
potential against non-perturbative effects. Indeed, given that an
anti-brane is typically repelled from the boundaries by membrane
instanton effects while being attracted to positively charged
boundaries by Coulomb forces, it seems likely that stabilization can
be achieved.  In addition, the non-holomorphic nature of the
gauge-kinetic function and its dependence on all brane moduli means
that non-perturbative potentials due to gaugino condensation will be
different from what one might have naively expected. Non-perturbative
potentials and moduli stabilization in the presence of anti-branes
will be explicitly studied in separate
publications~\cite{otherpapers}.

The plan of the paper is as follows. In the next section, we describe
our theory in five dimensions and present the associated
five-dimensional action of heterotic M-theory, including
anti-branes. Section~\ref{dw} discusses the various warping effects in
this five-dimensional theory within the context of a toy model and
explicitly presents the five-dimensional non-supersymmetric domain
wall solution. In Section~\ref{reduction}, we calculate the
four-dimensional bosonic effective theory to first order, that is, to
order $\kappa_{11}^{\frac{2}{3}}$, and discuss our results. Some
results at order $\kappa_{11}^{\frac{4}{3}}$, specifically the
complete corrections to the anti-brane potential and the gauge kinetic
functions at this order, are presented in Section~\ref{secondorder}.
A simple example of our results is provided in
Section~\ref{examplesection}. We conclude in Section~\ref{Conclusion}.
A number of technical Appendices explain the origin of the
five-dimensional theory in terms of the underlying 11-dimensional
one. In addition, they contain detailed technical results for the
five-dimensional gravitational warping which is needed in the
reduction to four dimensions.

\section{The Action}
\label{5dactionsection}

The theory we consider is the following. Start with a five-dimensional
description of heterotic
M-theory~\cite{Lukas:1998yy,Lukas:1998tt,Brandle:2001ts}, where six of
the dimensions of the underlying Ho\v{r}ava-Witten
theory~\cite{Horava:1996ma} have been compactified on any smooth
Calabi-Yau manifold $X$. We include an arbitrary number of M
five-branes which are parallel to the orbifold fixed planes and wrap
holomorphic curves in the Calabi-Yau space. In addition, we add a
single anti M five-brane which is associated with an anti-holomorphic
curve and is also taken to be parallel to the orbifold fixed
planes. The theory is easily generalized to an arbitrary number of
anti-branes, but we restrict ourselves to one such object for
simplicity. Most of the results which we derive are, in fact, valid
for arbitrary numbers of anti-branes, as will be discussed in more
detail in the text.  M five-branes are chosen parallel to the orbifold
planes so as to preserve ${\cal{N}}=1$ supersymmetry in the effective
theory. We have chosen this configuration for the anti-brane as well
because after all effects, both perturbative and non-perturbative,
have been included in the four-dimensional effective theory, we want
the vacua obtained to be maximally symmetric. Choosing the reduction
ansatz to include a slowly varying position modulus for the
anti-brane, which describes its displacement from some locus parallel
to the fixed planes, ensures that the regime of physical interest is
within the regime of validity of the effective theory. We choose the
anti-brane to wrap a holomorphic curve for two reasons. First, such a
two-cycle is volume minimizing in its homology class and so
constitutes a natural choice for a vacuum state. Second, such a choice
leads, upon integrating out the Calabi-Yau space, to a supersymmetric
theory in five dimensions. The existence of such structure will afford
us certain technical advantages in this and future work
\cite{otherpapers}. Of course, from the perspective of
five-dimensional heterotic M-theory, the (anti) M five-branes appear
as (anti) three-branes and we will refer to them as such in the
following.

\begin{figure}[ht]\centering 
\includegraphics[height=9cm,width=14cm, angle=0]{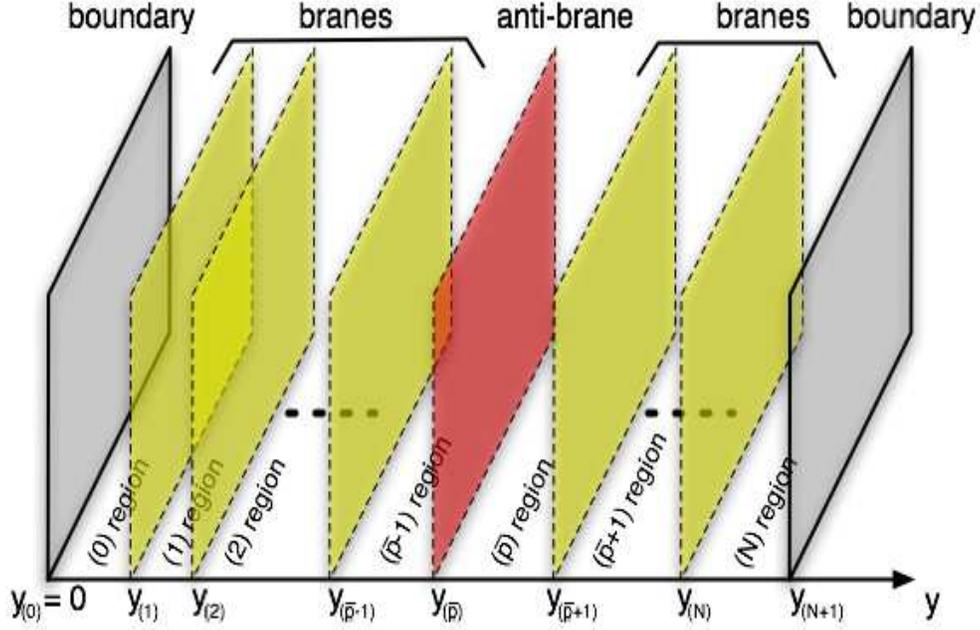}
        \caption{The brane configuration in five-dimensional heterotic M-theory}
\label{fig1}
\end{figure}

The general brane configuration is depicted in Fig. \ref{fig1}. 
The space-time
of five-dimensional heterotic M-theory consists of $S^1/{\mathbb Z}_2$ (
or, equivalently, an interval) times a
four-dimensional space-time with Minkowski signature. Five-dimensional
coordinates are denoted by $(x^\alpha )=(x^\mu ,y)$ where $y\in
[-\pi\rho,\pi\rho ]$ is the coordinate along the $S^1$ orbi-circle. The two
four-dimensional orbifold boundaries are then located at $y=0$ and 
$y=\pm\pi\rho$.
Between those boundaries we have a total of $N$ branes of which $N-1$
are three-branes and the remaining one is the anti three-brane.  We
will use indices $p,q,\dots =0,\dots , N+1$ to label all these
four-dimensional extended objects, where $p=0$ and $p=N+1$
correspond to the orbifold boundaries, $p=\bar{p}$ corresponds to the anti
three-brane and all other values of $p$ refer to three-branes.
Also note from Fig.~\ref{fig1} that the region between brane $p$
and brane $p+1$ is denoted by $(p)$, a notation that will
allow us to easily specify a field configuration in a specific part of
the interval. Since the sources on the branes and boundaries lead to
fields which are generally not smooth along the interval, this
notation for the various segments of the interval will be useful.
The world-volume coordinates of the $p$-th brane are denoted by $ \sigma^\mu_{(p)}$
and its embedding into five-dimensional space-time is given by
\begin{equation}
\label{embedcoord}
x^{\mu} = \sigma^{\mu} \;, \qquad y = y_{(p)}(\sigma^{\mu}_{(p)})\; .
\end{equation}
That is, the position of the $p$-th brane in the orbifold
direction\footnote{If we work in the ``upstairs'' picture, that is, with
  the full orbi-circle, we will have to include ``mirror branes'' at
  $y=-y_{(p)}$, where $p=1,\dots ,N$ in order to have a ${\mathbb
    Z}_2$ symmetric configuration.} is determined by $y_{(p)}$, where
$p=1,\dots , N$. Even though the orbifold boundaries are non-dynamical, it is
convenient to introduce trivial embedding coordinates $y=y_{(0)}=0$
and $y=y_{(N+1)}=\pi\rho$ for them. For ease of notation, we also
denote the embedding coordinate of the anti three-brane as
$\bar{y}=y_{({\bar p})}$ .  We will also frequently use
normalized orbifold coordinates $z=y/\pi\rho\in [0,1]$ and
$z_{(p)}=y_{(p)}/\pi\rho\in [0,1]$, where $p=0,\dots ,N+1$, to
simplify our notation.

Finally, we should briefly discuss the charges and tensions of the
orbifold boundaries and branes. For this purpose, it is useful to introduce an
integral basis $C^k$, where $k,l,\dots =1,\dots ,h^{1,1}(X)$, of the
second homology of the internal Calabi-Yau space $X$. Suppose that the
$p$-th M five-brane, where $p=1,\dots ,N$, wraps the cycle
$C^{(p)}$ given by
\begin{equation}
 C^{(p)}=\beta^{(p)}_kC^k\; .
\label{last1}
\end{equation}
Then the integer coefficients $(\beta^{(p)}_k)$ represent the charge vector
of the brane $p$. The charges on the orbifold boundaries are determined by
the second Chern classes $c_2(X)$, $c_2(V_{(0)})$ and $c_2(V_{(N+1)})$ of
the Calabi-Yau space $X$ and the two internal vector bundles $V_{(0)}$ and
$V_{(N+1)}$ on the boundaries, respectively.
More explicitly, we can define the charge vectors $(\beta^{(0)}_k)$
and $(\beta^{(N+1)}_k)$ of the two boundaries by
\begin{equation}
 c_2(V_{(0)})-\frac{1}{2}c_2(X)=\beta^{(0)}_kC^k\; ,\qquad
  c_2(V_{(N+1)})-\frac{1}{2}c_2(X)=\beta^{(N+1)}_kC^k\; .
\end{equation}
Heterotic anomaly cancellation dictates that 
\begin{equation}
\sum_{p=0}^{N+1}\beta^{(p)}_k=0\label{anomaly}
\end{equation}
for each $k=1,\dots ,h^{1,1}(X)$.  For the orbifold boundaries and the
three-branes the tensions $\tau^{(p)}_k$ are equal to the charges, so
we have $\tau^{(p)}_k=\beta^{(p)}_k\mbox{ for }p\neq\bar{p}$.  For the
anti three-brane, on the other hand, tensions and charges have
opposite signs, so $\tau^{({\bar p})}_k=-\beta^{({\bar p})}_k$. It
should be noted that the tension of the anti three-brane is positive.
In subsequent equations it will often be instructive to single out
terms related to the anti three-brane and, when doing so, we will
simply write the tension and charge of the anti three-brane as
$\bar{\tau}_k\equiv\tau_k^{({\bar p})}$ and
${\bar\beta}_k\equiv\beta_k^{({\bar p})}$. It is also useful to
introduce the step-functions
\begin{equation}
 \hat{\beta}_k(y)=\sum_{p,y_{(p)}<y}\beta_k^{(p)}
\end{equation}
which represent the sum of the charges to the left of a given point
$y$ in the interval.

\subsection{The Five-Dimensional Theory}
\label{5d}

Our starting point is the action of five-dimensional heterotic
M-theory~\cite{Lukas:1998yy,Lukas:1998tt,Brandle:2001ts}, obtained
from 11-dimensional Ho\v{r}ava-Witten theory~\cite{Horava:1996ma}
by compactifying on a Calabi-Yau manifold $X$ in the presence of both
M5 and anti-M5 branes. In this subsection, we will describe the field
content of this theory and then present the action itself. Many of the
detailed definitions of quantities which appear here can be found in
the Appendix.

We start by describing the bulk field content, focusing on
the bosonic degrees of freedom. In addition to the five-dimensional
metric $g_{\alpha\beta}$, the bulk fields consist of a K\"ahler sector,
labeled by indices $k,l,\dots = 1,\dots ,h^{1,1}(X)$ and a complex
structure sector, labeled by indices $a,b,\dots = 1,\dots ,h^{2,1}(X)$
or $A,B,\dots =0,\dots ,h^{2,1}(X)$.  In the K\"ahler sector, we have
$h^{1,1}(X)$ Abelian vector fields ${\cal A}_{\alpha}^k$ with field strengths
${\cal F}_{\alpha\beta}^k$ which descend from the M-theory three-form, a real
scalar field $V$ which measures the Calabi-Yau volume and $h^{1,1}(X)$
metric K\"ahler moduli $b^k$ which obey the condition $d_{i j k} b^i
b^j b^k =6$ and measure the relative size of the Calabi-Yau
two-cycles. In the complex structure sector, we have $h^{2,1}(X)$
metric complex structure moduli ${\mathfrak z}^a$ and $2(h_{2,1}+1)$
scalar fields $\xi^A$ and $\tilde{\xi}_B$ which descend from the
M-theory three-form. This three-form also gives rise to a
five-dimensional three-form $C_{\alpha \beta \gamma}$ and its field
strength $G_{\alpha\beta \gamma \delta}$. Of these bulk fields, $V$,
${\mathfrak z^a}$, $g_{\mu\nu}$, $g_{yy}$, $b^k$, $C_{\mu \nu y}$ and
${\cal A}^k_y$ are even under the ${\mathbb Z}_2$ action of the
$S^1/{\mathbb Z}_2$ orbifold, while $g_{\mu y}$, $\xi^A$,
$\tilde{\xi}_B$, ${\cal A}^k_{\mu}$, $C_{\mu \nu \gamma}$ are odd. 
The 11-dimensional origin of these fields is explained in Appendix~\ref{App:mod}.

The bulk theory is a $d=5$, ${\cal N}=1$ supergravity and, hence, the
various fields above should form the bosonic parts of five-dimensional
supermultiplets. The bosonic field content of the five-dimensional
supergravity multiplet consists of the five-dimensional metric
$g_{\alpha\beta}$ and an Abelian gauge field, which can be identified
as the linear combination $b_k{\cal A}_{\alpha}^k$. The remaining
$h^{1,1}(X)-1$ vectors, together with the $h^{1,1}(X)-1$ K\"ahler
moduli $b^k$, form the bosonic parts of $h^{1,1}(X)-1$ vector
multiplets. The remaining scalar fields, that is, the Calabi-Yau volume
modulus $V$, the dual of the three-form $C_{\alpha\beta\gamma}$, the
complex structure moduli ${\mathfrak z}^a$ and the axions $\xi^A$,
$\tilde{\xi}_B$ account for the bosonic parts of $h^{2,1}+1$
hypermultiplets, each of which contains four real scalars.

As usual, we have additional degrees of freedom which live on the
orbifold boundaries and branes. On the four-dimensional
orbifold boundaries, labeled by $p=0,N+1$, we have ${\cal N}=1$ gauge theories
with gauge fields $A_{(p) \mu}$ transforming in the adjoint of the
gauge groups ${\cal H}_{(p)}\subseteq E_8$ and gauge 
matter fields in ${\cal N}=1$
chiral multiplets with scalar components $C^{Ix}_{(p)}$. They
transform in representations of ${\cal H}_{(p)}$ which we shall denote
by $R_{(p)I}$, with $I,J,\dots$ labeling the different
representations and $x,y\dots$ the states within each representation.
More details on the origin and structure of the matter sector can be
found in Appendix~\ref{App:boundaries}.

The world-volume fields associated with the three-branes which descend
from wrapping an M5 brane on a holomorphic (or anti-holomorphic) curve
in the Calabi-Yau space are the following. The embedding coordinate (brane
position) $y_{(p)}$ together with the world volume scalar $s_{(p)}$
which descends from the two-form on the M five-brane world-volume,
pair together to form the bosonic content of an ${\cal N}=1$ chiral
multiplet, $(y_{(p)},s_{(p)})$. In addition, we have ${\cal N}=1$ gauge
multiplets with the associated field strengths denoted by
$E^{u}_{(p)}$. Here $u,v\dots =1,...,g_{(p)}$, where $g_{(p)}$ is the
genus of the curve $C^{(p)}$ wrapped by the $p$-th M5 brane. In
general, there will be additional chiral multiplets describing the
moduli space of the five brane curves and non-Abelian generalizations
of the gauge field degrees of freedom when M5 branes are
stacked. These are not vital to our discussion and we will not
explicitly take them into account. A similar selection
of four-dimensional fields appears on the anti-brane world volume.

\vspace{0.1cm}

Given this field content, the following is the bosonic part of the
five-dimensional action
describing Ho\v{r}ava-Witten theory~\cite{Horava:1996ma} compactified
on an arbitrary Calabi-Yau
manifold~\cite{Witten:1996mz,Lukas:1997fg,Lukas:1998yy,Lukas:1998tt,Brandle:2001ts}
in the presence of M5 and anti-M5 branes.
\bea \label{5daction}\nonumber S = - \frac{1}{2 \kappa_5^2} \int d^5 x
\sqrt{-g} \left[ \frac{1}{2} R + \frac{1}{4} G_{kl}(b) \partial
  b^k \partial b^l + \frac{1}{2} G_{kl}(b) {\cal F}^k_{\alpha \beta}
  {\cal F}^{l \alpha \beta} + \frac{1}{4} V^{-2} (\partial V)^2 +
  \lambda (d_{i j k} b^i b^j b^k -6)\right. \\ \nonumber \left. +
  \frac{1}{4} {\cal K}_{a \bar{b}}({\mathfrak z})\partial {\mathfrak
    z}^a
  \partial \bar{{\mathfrak z}}^{\bar{b}} - V^{-1} (\tilde{\cal
    X}_{A\alpha} - \bar{M}_{AB} ({\mathfrak z}){\cal X}_{\alpha}^B) (
  [ \textnormal{Im}{(M ({\mathfrak z}))}]^{-1})^{AC} (\tilde{\cal
    X}_C^\alpha - M_{CD} ({\mathfrak z}){\cal X}^{D\alpha}) \right.\\
\nonumber \left. +\frac{1}{4!} V^2 G_{\alpha \beta \gamma \delta}
  G^{\alpha \beta \gamma \delta} + m^2V^{-2} G^{kl} (b) \hat{\beta}_k
  \hat{\beta}_l \right] \\ \nonumber -\frac{1}{2 \kappa_5^2} \int
\left( \frac{2}{3} d_{klm} {\cal A}^k \wedge {\cal F}^l \wedge {\cal
    F}^m + 2 G \wedge (( \xi^A \tilde{\cal X}_A - \tilde{\xi}_A{\cal
    X}^A) - 2 m\hat{\beta}_k {\cal A}^k) \right)\eea

\bea \label{action} - \int d^5 x \; \delta (y)
 \sqrt{- h_{(0)} }\,\left[ \frac{m}{\kappa_5^2} V^{-1} b^k \tau^{(0)}_k +
  \frac{1}{16 \pi \alpha_{\textnormal{GUT}}} V \rm{tr}(F_{(0)}^2)  +
  G_{(0)IJ} D_{\mu} C^{Ix}_{(0)} D^{\mu} \bar{C}^J_{(0) x}
 \right. \\ \left. \nonumber + V^{-1} G^{IJ}_{(0)} \frac{\partial
    W_{(0)}}{\partial C^{Ix}_{(0)}} \frac{\partial
    \bar{W}_{(0)}}{\partial \bar{C}^J_{(0) x}} + {\rm tr}(D_{(0)}^2) \right] \eea

\bea \nonumber -  \int d^5 x \; \delta (y - \pi
\rho) \sqrt{- h_{(N+1)} }\,\left[\frac{m}{\kappa_5^2} V^{-1} b^k \tau^{(N+1)}_k +
  \frac{1}{16 \pi \alpha_{\textnormal{GUT}}} V {\rm tr}(F_{(N+1)}^2)
  \right. \\ \left. \nonumber + G_{(N+1)IJ}
  D_{\mu} C^{Ix}_{(N+1)} D^{\mu} \bar{C}^J_{(N+1) x} + V^{-1} G^{IJ}_{(N+1)} \frac{\partial W_{(N+1)}}{\partial
    C^{Ix}_{(N+1)}} \frac{\partial \bar{W}_{(N+1)}}{\partial
    \bar{C}^J_{(N+1) x}} + {\rm tr}(D_{(N+1)}^2) \right]
\eea

\bea \nonumber - \frac{1}{2 \kappa_5^2} \int d^5 x
\;\left\{\sum_{p=1}^{N} (\delta (y- y_{(p)})+\delta (y+y_{(p)}))
  \sqrt{-h_{(p)}}\,\left[ mV^{-1} \tau^{(p)}_k b^k + \frac{2 m (
      n_{(p)}^k \tau^{(p)}_k)^2}{V (\tau^{(p)}_l b^l)} j_{{(p)} \mu}
    j^{\mu}_{(p)} \right. \right. \\ \left. \left. \nonumber + [
    \textnormal{Im}{\Pi}]_{{(p)} uw} E^u_{{(p)} \mu \nu} E^{w \mu
      \nu}_{(p)} \right] - 4 m \hat{C}_{(p)} \wedge \tau^{(p)}_k
  d(n^k_{(p)} s_{(p)}) - 2 [ \textnormal{Re}{ \Pi}]_{(p)uw} E^u_{(p)}
  \wedge E^w_{(p)} \right\} \eea Let us briefly discuss the various
quantities in this action. In the previous section, we defined
the charges $\beta^{(p)}_k$, the tensions $\tau^{(p)}_k$ and the
charge step-functions $\hat{\beta}_k$.  To introduce the remaining
objects, we start with the bulk theory, the first part of the above
action. Of course, $\kappa_5$ is the five-dimensional Planck constant,
related to its 11-dimensional counterpart $\kappa_{11}$ by
$\kappa_5^2=\kappa_{11}^2/v$ where $v$ is the Calabi-Yau reference
volume.  The constant $m$ is given by
\begin{equation}
\label{add1}
 m=\frac{2 \pi}{v^{\frac{2}{3}}}
 \left( \frac{\kappa_{11}}{4 \pi} \right)^{\frac{2}{3}}
\end{equation}
and represents a reference mass scale of the Calabi-Yau space. The
quantity $\lambda$ is a Lagrange multiplier enforcing the constraint
on the $b^k$ moduli.  The K\"ahler and complex structure moduli
metrics $G_{kl}$ and ${\cal K}_{a\bar{b}}$ are defined in
Appendix~\ref{App:mod}.  A definition of the Calabi-Yau intersection
numbers $d_{ijk}$ and the special geometry quantity $M_{AB}$ can also
be found in this Appendix.  The various bulk form field strengths are
defined in the usual way as $G=dC$, ${\cal F}^k=d{\cal A}^k$ and
${\cal X_A}=d\xi_A$, $\tilde{\cal X}_A=d\tilde{\xi}_A$ away from the
boundaries, but are subject to boundary source terms specified by the
relations \bea \label{BI1} (dG)_{y \mu \nu \gamma \rho} &=& -4
\kappa_5^2 ( J^{(0)}_{4\mu \nu \gamma \rho} \delta (y) +
J^{(N+1)}_{4\mu\nu\gamma\rho} \delta (y- \pi \rho )) \\ \label{BI2} (d
{\cal F}^k)_{y \mu \nu} &=& -4 \kappa^2_5 ( J^{(0) k}_{2\mu \nu}
\delta (y) + J^{(N+1) k}_{2\mu \nu} \delta (y - \pi \rho))
\\ \label{BI3} (d {\cal X}^A {\cal G}_A - d \tilde{{\cal X}}_B {\cal
  Z}^B)_{y \mu} &=& -4 \kappa^2_5 ( J^{(0)}_{1\mu} \delta (y) +
J^{(N+1)}_{1\mu} \delta (y - \pi \rho)) \eea where \bea J^{(p)}_{4\mu
  \nu \gamma \rho} &=& \frac{1}{16 \pi \alpha_{\textnormal{ GUT}}}
\textnormal{tr} (F_{(p)} \wedge F_{(p)})_{\mu \nu \gamma \rho}
\\ \label{mattercurrent} J^{(p)k}_{2\mu \nu} &=& -i \sum_{I,J}
\Gamma^k_{ (p)IJ} ( D_{\mu} C_{(p)}^{Ix} D_{\nu} \bar{C}_{(p)x}^J -
D_{\mu}
\bar{C}_{(p)x}^I D_{\nu} C_{(p)}^{Jx} ) \label{J2}\\
J^{(p)}_{1\mu} &=& \frac{e^{-{\cal K}}}{2 V}\sum_{I,J,K} \lambda_{IJK}
f_{xyz}^{(IJK)} C_{(p)}^{Ix} C_{(p)}^{Jy}D_{\mu} C_{(p)}^{Kz} \eea for
$p=0,N+1$. The various matter field objects in these sources are
defined in Appendix~\ref{App:boundaries}. One important observation
from these Bianchi identities is that the three-branes and anti
three-branes do not contribute any source terms. This fact will be
crucial in our later analysis.  The second and third parts of the
above action are the theories on the two orbifold boundaries
respectively. They are written in terms of the matter field K\"ahler
metrics $G_{(p)MN}$, the matter field superpotentials $W_{(p)}$ and
the D-terms $D_{(p)}$. Definitions for these quantities can also be
found in Appendix~\ref{App:boundaries}.  The (reference) gauge coupling constant
$\alpha_{\textnormal{GUT}}$ is given by
$\alpha_{\textnormal{GUT}}=(4\pi\kappa_{11}^2)^{2/3}/v$.

We move on to discuss the three-brane world volume theories, the last
part of the above action. The quantities
$n_{(p)}^k=\beta^{(p)}_k/\sum_{l=1}^N{\beta^{(p) 2}_l}$ are a normalized
version of the three-brane charges and the axionic currents
$j_{(p)\mu}$ are defined by
 \bea
j_{(p)\mu} = \frac{\beta^{(p)}_k}{n_{(p)}^l \beta^{(p)}_l} ( d(n^k_{(p)} s_{(p)})
                - {\hat{\cal A}}^k_{(p)} )_{\mu} \; ,
\eea
where $\hat{C}_{(p)}$ and ${\hat{\cal A}}^k_{(p)}$ denote the
pull-backs of the bulk forms $C$ and ${\cal A}^k$ to the $p$-th
brane. The gauge kinetic functions $\Pi_{(p)uv}$ of
the three-brane gauge fields are defined in Appendix~\ref{App:5brane}.

Finally, we need to mention that the induced metrics $h_{(p)\mu\nu}$ on the
orbifold boundaries and branes are explicitly given by
\bea
\label{inducedmetric}
h_{(p)\mu \nu} = g_{\mu \nu} + g_{\mu y} \partial_{\nu} y_{(p)} + g_{y
  \nu} \partial_{\mu} y_{(p)} + g_{yy} \partial_{\mu}
y_{(p)} \partial_{\nu} y_{(p)}\; , \eea where the
embedding~\eqref{embedcoord} has been used. Recall that the boundaries
are non-dynamical with associated static embeddings $y_{(0)}=0$ and
$y_{(N+1)}=\pi\rho$. Hence, the induced boundary metrics $h_{(0)}$ and
$h_{(N+1)}$ are simply equal to $g_{\mu\nu}$, the four-dimensional
part of the bulk space-time metric. The action described in this
section must be supplemented by the usual Gibbons-Hawking boundary
terms. A careful analysis reveals that form-field boundary terms are
not required in this case.

Having described the five-dimensional theory, our starting
point, we proceed in the next section to discuss the
appropriate reduction ansatz in the presence of
anti-branes. The dimensional reduction to four-dimensions will
be performed in section~\ref{reduction}.

\section{The Five-Dimensional Domain Wall with Anti-Branes}
\label{dw}
In this section, we illustrate the main features of the
five-dimensional reduction ansatz in the context of a simple scalar
field toy model.  We will then explicitly work out the essential part
of this reduction ansatz, the five-dimensional non-supersymmetric
domain wall. This is a generalization of the BPS domain wall solution
of Ref.~\cite{Lukas:1998yy,Lukas:1998tt} and includes the
back-reaction effects of the anti three-brane.

The key new point for us will be to discover how the back-reaction on
the bulk fields due to the presence of the branes and, in particular,
the anti-brane is taken into account in the reduction ansatz. While
this is technically complicated for five-dimensional heterotic
M-theory, the basic ideas can be explained in a simple setting. Before
dealing with the full problem, we will, therefore, discuss a scalar
field toy model~\cite{Lukas:1998ew} to illustrate the key features
involved. The structure of space-time and branes for this model is
precisely as described above and illustrated in Fig.~\ref{fig1}. The
action is given by
\begin{equation}
  S\sim\int d^5x\,\left[\partial_\alpha\Phi\partial^\alpha\Phi 
    -\delta (y)\,S_{(0)}\Phi
    -\delta (y-\pi\rho )\,S_{(N+1)}\Phi
    -\sum_{p=1}^N(\delta (y-y_{(p)})+\delta (y+y_{(p)}))\,S_{(p)}\Phi\right]\; ,
\end{equation}
where $S_{(p)}$ are sources on the boundaries and branes (which can
depend on other fields) and $\Phi$ is a ${\mathbb Z}_2$ even scalar
field. What we want to discuss in this model is the warped background
solution which arises due to the presence of the source terms and the
four-dimensional effective theory associated with it. To this end, it
is useful to split the scalar field as $\Phi = \phi +\phi_0$, where
$\phi_0$ is a function of the four-dimensional coordinates only and is
the quantity that will become the modulus associated with this degree
of freedom in the four-dimensional effective theory. On the other
hand, $\phi$ represents a function of all five coordinates and
contains the warping of the background due to the presence of sources
terms. To uniquely define this splitting of $\Phi$, we require that
the orbifold average $<\phi>$ of $\phi$ vanishes. This condition
implies a specific choice of coordinates on field space in the
resulting four-dimensional effective theory. This choice is
particularly useful in finding a clean form for the resulting action,
as we will see explicitly throughout this paper.

The field equation for $\Phi$, valid in each bulk region indicated
in Fig.~\ref{fig1}, then reads
\begin{equation}
\label{schematiceom}
\Box_4 \phi_0 + \Box_4 \phi + D_y^2 \phi = 0\; .
\end{equation} 
In addition, $\Phi$ is subject to boundary conditions
at the edge of each region due to the presence of the sources. For the
two orbifold boundaries, these take the form
\begin{equation}
\label{schematicbc}
D_y \phi|_{y=0} = - S_{(0)}\; ,\qquad D_y \phi|_{y=\pi\rho} = + S_{(N+1)}
\end{equation}
while, for the branes, we have
\begin{equation}
\label{schematicjump}
- D_y \phi|_{y=y_{(p)}+} + D_y \phi|_{y=y_{(p)}-} = S_{(p)}\; ,
\end{equation}
where $p=1,\dots , N$.
The subscript ``$y=y_{(p)}+$'' (``$y=y_{(p)}-$'') indicates that the relevant quantity should
be evaluated approaching the $p$-th brane from the right (left).

We can now take an average of the equation of motion \eqref{schematiceom} over
the orbifold. Using $<\phi>=0$ and the ``boundary conditions''
(\ref{schematicbc}) and (\ref{schematicjump}), we obtain
\begin{equation}
\label{schematicaverage}
\Box_4 \phi_0 + \sum_p S_{(p)} =0 \; .
\end{equation}
This relation may then be used to eliminate $\phi_0$ in
\eqref{schematiceom} to obtain an equation purely for the warping
\begin{equation}
\label{schematicwarping}
\Box_4 \phi + D^2_y \phi = \sum_p S_{(p)}.
\end{equation}
To pursue the analysis further, we need to know something about the
various approximations, and associated expansions, which are made in
deriving four-dimensional heterotic M-theory, some of which have
already been implicit in our analysis. Two expansions in particular
are of central importance at this point. The first of these is simply
the usual expansion in four-dimensional derivatives which is always
made in defining such an effective theory; in other words, the 
four-dimensional fields are assumed to be slowly varying relative to the
structure in the internal dimensions. The second expansion which we
need is in terms of a small parameter $\epsilon_S$, which controls
the size of the source terms. We will meet this
quantity explicitly soon, so let us just state this to be true for
now. The zero mode $\phi_0$ is a quantity independent of the warping and
is, therefore, zeroth order in the $\epsilon_S$ expansion. By contrast,
$\phi$ is precisely the quantity which presents the warping
and so is first order $\epsilon_S$. Looking
at \eqref{schematicwarping} which determines the warping, we see that
the first term is both first order in $\epsilon_S$ and second order in
four-dimensional derivatives whereas the remaining terms 
are simply first order
in $\epsilon_S$. We may therefore, in a controlled approximation,
ignore the first term in~\eqref{schematicwarping}. This results in the following equation for the
warping  \begin{equation}
\label{schematicfinal}
D_y^2 \phi = \sum_p S_{(p)}.
\end{equation} 
Thus, in the end, we need to solve the system of bulk equations and
boundary conditions given by~\eqref{schematicbc}, \eqref{schematicjump}
and \eqref{schematicfinal}. Before
moving on to the full calculation, we qualitatively
describe such an analysis in various cases by transferring the insight
from the above toy example to heterotic M-theory. We start with
heterotic M-theory in the absence of anti-branes and proceed to
a discussion of the back-reaction of such objects when they are included
in the vacuum.

\vspace{0.1cm}

\paragraph*{\bf a) Zeroth order in sources}
When all of the sources are set to zero, the warping
equation~\eqref{schematicfinal} becomes simply $D^2_y \phi =0$. Since
$\phi$ must be continuous around the orbifold this, in combination
with the condition $<\phi>=0$, results in $\phi = 0$. For five-dimensional
heterotic M-theory, this implies that the zeroth order vacuum is simply
five-dimensional Minkowski space.

\vspace{0.1cm}

\paragraph*{\bf b) The standard heterotic vacuum}
In the case of a heterotic vacuum involving only orbifold fixed planes
and three-branes (but no anti three-branes), the sources $S_{(p)}$ are
represented by the tension terms of the branes and boundaries and they
obey a particularly useful relation. Since the objects involved are
all BPS, their tensions are equal to their charges. The charges, on
the other hand, have to sum to zero as a result of the heterotic
anomaly cancellation condition~\eqref{anomaly}.  Thus, we have for such
compactifications that $\sum_p S_{(p)} =0$ and, as in the previous
case, the equation for the warping \eqref{schematicfinal} reduces to
$D^2_{y} \phi =0$. The boundary conditions
\eqref{schematicbc}, \eqref{schematicjump} are no longer trivial however.
Thus, the only solution for $\phi$ is a function linear 
in each region of Fig.~\ref{fig1}, with kinks at the boundary and brane
positions. The slope in each region is
chosen such that equations \eqref{schematicbc}, \eqref{schematicjump} are
obeyed. The fact that the source terms sum to zero ensures that
we indeed have a globally well-defined solution. This situation
corresponds to the ``Universe as a Domain Wall'' vacuum of heterotic
M-theory~\cite{Lukas:1998yy,Lukas:1998tt}.

\vspace{0.1cm}

\paragraph*{\bf c) Matter fluctuation induced warping in the case without anti-branes}
We have just seen that the vacuum of heterotic M-theory is linearly warped if
only tension sources of BPS objects are considered. However, even in the
absence of anti-branes this is not the full story~\cite{Lukas:1997fg,Lukas:1998ew}.
For example, fluctuations of matter fields on the orbifold fixed planes
lead to additional source terms. Since fields on 
the various orbifold boundaries and
branes fluctuate independently, these sources do not, in general, obey
the sum rule $\sum_pS_{(p)}=0$. As a result, we obtain warping quadratic in
the orbifold coordinate $y$. This is typical of heterotic M-theory: 
any change to
the source terms away from the case of ``pure tension''
results in quadratic contributions to the warping.

\vspace{0.1cm}

\paragraph*{\bf d) The back-reaction of anti-branes in heterotic M-theory}

We are now in a position to understand how the presence of an
anti-brane in the bulk of heterotic M-theory changes the warping. In
other words, we can now see how to calculate the back-reaction due to
the presence of the anti-brane. The relevant source terms - the
tensions of the orbifold boundaries, the three-branes and the anti
three-branes - are very similar to those for the usual BPS situation
described in b). While the charges of these objects still add up to
zero, the same can no longer be said for the tensions since the anti
three-brane tension is minus its charge.  Hence, the bulk equation
\eqref{schematicfinal} contains a non-vanishing source term
$\sum_pS_{(p)}\neq 0$ due to the tension terms. It is clear from the
discussion in c) that this implies a quadratic term in the warping.
More precisely, using the usual linear warping we can ensure that all
boundary conditions, except the one at the right boundary $y=\pi\rho$,
are satisfied.  Since the sum of the sources no longer vanishes, the
linear warping does not match the final boundary condition at
$y=\pi\rho$. We therefore add a quadratic piece to the warping. This
does not change the kink structure of the solution across the branes,
but does allow one to satisfy the boundary condition at $y=\pi\rho$.
In addition, when we examine the bulk equation \eqref{schematicfinal},
we find that this additional quadratic warping is precisely what is
required to balance the source term now present on the right-hand
side.

\vskip 0.4cm

With the insight from this toy model, we now proceed to analyze full
five-dimensional heterotic M-theory. In the remainder of this section,
we focus on the warping caused by the tension terms. This will lead us
to a generalization of the heterotic domain wall vacuum, valid in the
presence of an anti-brane. The additional warping due to fluctuations
of localized fields is presented in Appendix~\ref{App:Bianchi}.

The only bulk fields involved in the generalized domain wall solution
are the ones which couple to the tension terms in the
action~\eqref{5daction}. They are the metric $g_{\alpha\beta}$, the
volume modulus $V$ and the K\"ahler moduli $b^k$. We start with the
usual metric ansatz involving the four dimensional metric $g_{4 \mu
  \nu}$.
\begin{eqnarray}
  ds^2_5 &=& a^2(y,x^{\mu}) g_{4 \mu \nu}d x^{\mu} d x^{\nu} + b^2(y,x^{\mu}) d y^2\label{metric}\\
  V&=&V(y,x^{\mu})\\
  b^k&=&b^k(y,x^{\mu})\; .
\end{eqnarray}
In writing our result, it is useful to introduce the
following function which encodes the standard linear warping
of heterotic M-theory and averages to zero over the orbifold
\begin{equation}
h_{(p)k}(z) = \sum^{p}_{q=0} \tau^{(q)}_k (z-z_{(q)}) - \frac{1}{2}
\sum_{q=0}^{N+1} \tau^{(q)}_k z_{(q)}(z_{(q)} - 2) - \delta_k\; ,
\end{equation}
where we have defined the difference of anti three-brane tension and charge as
\begin{equation}
 \delta_k=\frac{1}{2}(\bar{\tau}_k-\bar{\beta}_k)=-\bar{\beta}_k\; . 
\label{delta}
\end{equation}
In the following, whenever we want to consider the supersymmetric limit
of our results, we can ``switch off'' the effect of the anti three-brane by formally 
setting $\delta^k\rightarrow 0$. Recall that the sub-script ``$(p)$'' indicates
that the domain of the function $h_{(p)k}$ is 
$z\in [z_{(p)},z_{(p+1)}]$, that is, the
region to the right of the $p$-th brane. Using the Einstein
equation and the equations of motion for $V$ and $b^k$ derived from
the action~\eqref{action}, together with the above ansatz, gives the
following solution for the warping
\begin{eqnarray}
 \frac{a_{(p)}}{a_0}&=&1-\epsilon_0\frac{b_0}{3V_0}b_0^k\left[ h_{(p)k}-
          \delta_k\left( z^2-\frac{1}{3}\right)\right]\label{a}\\
 \frac{V_{(p)}}{V_0}&=&1-2\epsilon_0\frac{b_0}{V_0}b_0^k\left[ h_{(p)k}-\delta_k
                      \left( z^2-\frac{1}{3}\right)\right]\label{V}\\
 b_{(p)}^k&=&b_0^k+2\epsilon_0\frac{b_0}{V_0}\left[\left(h_{(p)}^k-\frac{1}{3}h_{(p)l}b_0^kb_0^l
            \right)-\left( \delta^k-\frac{1}{3}\delta_lb_0^kb_0^l
           \right)\left( z^2-\frac{1}{3}\right)\right].\label{bk}
\end{eqnarray}
We remind the reader of the relationship $z=y/{\pi \rho}$. In these
expressions, $a_0$, $b_0$, $V_0$ and $b_0^k$ are four-dimensional
moduli fields. Note that due to our convention of zero average
warping, these moduli are precisely the orbifold average of the
corresponding five-dimensional fields. For example $V_0=<V>$ is the
orbifold average of the Calabi-Yau volume and $b_0=<b>$ is the average
orbifold radius. We observe that the warping in the above solution is
indeed proportional to the strong-coupling expansion parameter
$\epsilon_S$, defined by
\begin{equation}
\label{esdef}
 \epsilon_S =\epsilon_0\frac{b_0}{V_0}\; ,\qquad \epsilon_0=\pi\rho m
\end{equation}
as promised. Note using~\eqref{add1} that $\epsilon_S \propto
\kappa_{11}^{2/3}$.  Our result is valid as long as $\epsilon_S\ll 1$,
since we have neglected warping terms of order $\epsilon_S^2$ and
higher.  The structure of the ${\cal O}(\epsilon_S)$ warping terms is
as qualitatively described earlier. In the supersymmetric limit,
$\delta^k\rightarrow 0$, we recover the linear warping of the BPS
domain wall encoded in the functions $h_{(p)k}$.  On the other hand,
the terms proportional to $\delta^k$, which are caused by the presence
of the anti three-brane, represent quadratic warping. The observant
reader will note that we have not given an expression for the warping
of the metric coefficient $b$. This is because this $y$ dependence amounts to a coordinate
choice and, as such, is not needed in the calculation of the four
dimensional effective action.

\begin{figure}[ht]\centering 
\includegraphics[height=9cm,width=12cm, angle=0]{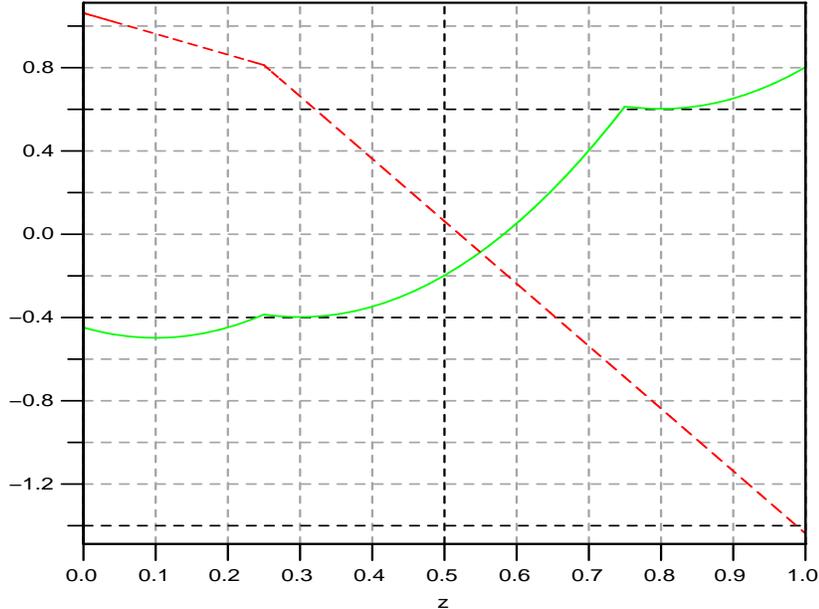}
\caption{Warping of the Calabi-Yau volume modulus V (bracket on the
  RHS of \eqref{V}), assuming $h^{1,1}(X)=1$. The dashed, red
  curve corresponds to a BPS configuration with one brane at $z=1/4$
  and charge vector $(\beta^{(p)})=(1,2,-3)$.  The solid, green curve
  describes a situation with one brane at $z=1/4$, one anti-brane at
  $z=3/4$ and charge vector $(\beta^{(p)})=(1,2,-5,2)$. It can be
  obtained from the previous BPS configuration by ``pulling'' an
  anti-brane with charge $-5$ off the boundary at $z=1$.}
\label{fig2}
\end{figure}

A specific example of the warping of the Calabi-Yau volume modulus
$V$ is plotted in Fig.~\ref{fig2}.  This example shows that the
presence of an anti-brane can change the warping substantially. As is
clear from the action~\eqref{5daction}, the volume $V$ at the
boundaries $z=0,1$ determines the value of the boundary gauge
couplings.  In particular, the BPS configuration in Fig.~\ref{fig2}
(dashed, red curve) corresponds to weak coupling at $z=0$ and strong
coupling at $z=1$. As is evident from the Figure, this behavior can
be reversed in the presence of the anti-brane (solid, green curve).
\section{Heterotic M-Theory with Anti-Branes in Four Dimensions}
\label{reduction}

In this section, we construct the four-dimensional effective theory
describing heterotic M-theory when anti-branes are present in the
vacuum, focusing on terms that are zeroth and first order in
$\kappa_{11}^{\frac{2}{3}}$.  In the next section, we will present
some terms of order $\kappa_{11}^{\frac{4}{3}}$ which can be reliably
calculated, concentrating on those which will be important in
applications to moduli stabilization.

\subsection{The Reduction Ansatz and Zero Modes}

We start by discussing the zero modes and the reduction ansatz used in
the rest of the paper. The essential part of the ansatz is the
non-supersymmetric domain wall, introduced in the previous section.
Let us focus on this part of the ansatz for now. We will discuss
additional terms due to fluctuations of localized fields later.  For
the five-dimensional metric, recall \eqref{metric},
\begin{equation}
 ds_5^2=a^2(y)g_{4\mu\nu}dx^\mu dx^\nu +b^2(y) dy^2\; ,
\end{equation}
where the warping $a(y)$ is given in \eqref{a}. The domain wall
also involves non-trivial warping for the volume modulus $V$ and the
K\"ahler moduli $b^k$, as given in~\eqref{V} and \eqref{bk}. In
addition to the four-dimensional metric, the complete solution
contains the moduli $V_0$, $a_0$, $b_0$ and $b_0^k$, which are
four-dimensional fields. Recall that $V_0$ is the orbifold average of
the Calabi-Yau volume and $b_0$ is the average radius of the orbifold
and we will frequently write
\begin{equation}
 b_0=e^{\beta}\; ,\qquad V_0=e^\phi\; .
\label{last2}
\end{equation}
The modulus $a_0$ is really the four-dimensional scale factor and is
redundant given that we have already introduced a full
four-dimensional metric into our ansatz.  However, it can be
conveniently used to bring the four-dimensional Einstein-Hilbert term
into canonical form, which is achieved by fixing
$a_0^2=1/b_0$. Further zero modes arise from the ${\mathbb Z}_2$ even
components of the other bulk fields.  Specifically, these are the zero
modes of the complex structure moduli ${\mathfrak z}^a$, the axions
$\chi^k={\cal A}^k_y$ and the two-form $B_{\mu\nu}=C_{y\mu\nu}$ with
field strength $H_{\mu\nu\rho}=G_{y\mu\nu\rho}$, which we dualize to a
scalar $\sigma$, as usual.  Note, in particular, that the fields
$\xi^A$ and $\tilde{\xi}_A$ are ${\mathbb Z}_2$ odd and do not give
rise to zero modes. However, the $y$-components ${\cal X}_y$ and
$\tilde{\cal X}_y$ of their field strengths are ${\mathbb Z}_2$ even
and having these components non-zero corresponds to switching on
$H$-flux from a 10-dimensional heterotic view point. Flux in the
context of heterotic models with anti-branes will be considered in
Ref.~\cite{otherpapers} but, for the purposes of this paper, we simply
set ${\cal X}_y$ and $\tilde{\cal X}_y$ to zero. In the following, we
will drop the ``$0$''sub-script from the four-dimensional fields for
ease of notation, for example, writing $b^k_0$ as $b^k$.

Fields localized on the orbifold boundaries and the branes are, of
course, already four-dimensional and are simply retained in the
four-dimensional effective theory.  In particular, for the
four-dimensional position moduli of the branes we will use the
normalized fields $z_{(p)}\in [0,1]$ and, in particular, $\bar{z}$ for
the anti three-brane. It is useful to normalize the associated brane
axion fields $s_{(p)}$ accordingly by defining
$\nu_{(p)}=s_{(p)}/\pi\rho$.  We also have the pull-backs of bulk
fields appearing on the boundaries and branes. These have to be
computed by applying the embedding~\eqref{embedcoord} to the reduction
ansatz for the bulk fields. For example, for the induced metrics on
the branes we have \bea \sqrt{-h_{(p)}} = a^4 \sqrt{-g_4} \left(1 +
  \frac{b^2}{2 a^2} \partial_{\mu} y_{(p)} \partial^{\mu} y_{(p)}
\right)\; .  \eea As discussed earlier, fluctuations of localized
fields give rise to additional warping contributions of the bulk
fields. These must be integrated out to arrive at the correct
effective action~\cite{Lukas:1997fg,Lukas:1998ew}. It turns out that,
to order $\kappa_{11}^{\frac{2}{3}}$, these effects are only relevant
for the bulk anti-symmetric tensor fields whose warping is induced by
the Bianchi identities~\eqref{BI1}--\eqref{BI3}. The solutions to
these Bianchi identities (and the associated equations of motion) in
the presence of fluctuating boundary fields are discussed in
Appendix~\ref{App:Bianchi}.

We shall present our results, specified order by order in the
$\kappa_{11}^{\frac{2}{3}}$ expansion parameter of Ho\v{r}ava-Witten
theory. This parameter
determines the computational complexity involved in obtaining the
relevant parts of the action. Thus the discussion of how to obtain the
second order pieces below is more involved than that describing the
first order results. It is useful, therefore, to separate them.
However, when applying these results the reader should keep in mind
that the correct expansion parameter to consider in four dimensions is
$\epsilon_S$. Thus all terms up to a given order in this parameter
should be kept. In particular, when working to first order in
$\epsilon_S$ the threshold corrections to the gauge kinetic functions
and matter field terms, which are second order in
$\kappa_{11}^{\frac{2}{3}}$, must be included.

There is a simple way in which the order in $\epsilon_S$ or
$\kappa_{11}^{\frac{2}{3}}$ of any term appearing in the following can
be ascertained.
\begin{itemize}
\item The order in $\epsilon_S$ of any term can be
  found by counting the number of powers of $\epsilon_0$ which are
  present.
\item The order in $\kappa_{11}^{\frac{2}{3}}$ of any term can 
be determined by counting the number of powers of
  $\epsilon_0$ which are present and then adding one power if the term
  contains boundary gauge or matter fields. This rule
  is a consequence of a rescaling which must be performed to obtain
  standard kinetic terms for these fields.
\end{itemize}

Having forewarned the reader of this subtlety, we now present our
results.

\subsection{Zeroth and First Order Result for the Four-Dimensional
  Effective Theory}
\label{01order}

Collecting various terms, we find the following bosonic
four-dimensional effective action of heterotic M-theory in the
presence of an arbitrary number of M five-branes and a single anti M
five-brane. This result is valid up to first order in
$\kappa_{11}^{\frac{2}{3}}$ and second order in four-dimensional
derivatives. We find that
\begin{equation}
S=S_{\delta^{0}}^{(0,1)}+S_{\delta^{1}}^{(1)},
\label{z1}
\end{equation}
where the subscript indicates that $S_{\delta^{0}}^{(0,1)}$ and $S_{\delta^{1}}^{(1)}$ 
contain terms independent of the parameters $\delta_{k}$ and linear in $\delta_{k}$ 
respectively. Similarly, the superscript implies that $S_{\delta^{0}}^{(0,1)}$ has both 
zeroth and first order terms in $\kappa_{11}^{\frac{2}{3}}$ whereas all terms in
$S_{\delta^{1}}^{(1)}$ are of order $\kappa_{11}^{\frac{2}{3}}$.
The first action in \eqref{z1} is given by
\begin{equation}
S_{\delta^{0}}^{(0,1)}= S_4^{\textnormal{moduli}}+ S_4^{\textnormal{gauge}}+
S_4^{\textnormal{matter}},
\label{z2}
\end{equation}
with
\bea \nonumber S_4^{\textnormal{moduli}} &=& -\frac{1}{2
  \kappa^2_4} \int d^4x\sqrt{-g_4}\, \left[ \frac{ 1}{2}R_4 +
  \frac{3}{4} ( \partial \beta)^2 +  \frac{1}{4} (\partial \phi)^2 +
 \frac{1}{4}e^{-2\phi} (\partial \sigma)^2+
 \frac{1}{4} G_{kl} \partial b^k \partial b^l + e^{-2 \beta} G_{kl}
 \partial \chi^k \partial \chi^l \right. \\  \label{resultmoduli}
 && \left. + \frac{1}{4}{\cal K}_{a
    \bar{b}}({\mathfrak z})
  \partial {\mathfrak z}^a \partial \bar{{\mathfrak z}}^{\bar{b}}+
  2\epsilon_0\sum_{p=1}^{N}
  \tau^{(p)}_kz_{(p)}e^{-2\phi}\partial\sigma\partial(n_{(p)}^k
  \nu_{(p)}) +
  \frac{\epsilon_0}{2}\sum_{p=1}^Nb^k\tau^{(p)}_ke^{\beta-\phi}
  (\partial z_{(p)})^2 \right. \\
\nonumber && \left. + 2\epsilon_0\sum_{p=1}^{N}
  \frac{\tau^{(p)}_l\tau^{(p)}_k}{\tau^{(p)}_mb^m}e^{-\phi -\beta}
  \left(\chi^l \chi^k (\partial z_{(p)})^2 - 2 \chi^k \partial
    (n_{(p)}^l\nu_{(p)}) \partial z_{(p)} +\partial (n_{(p)}^k
    \nu_{(p)}) \partial (n^l_{(p)} \nu_{(p)})\right) \right. \\
\nonumber && \qquad\qquad\qquad\qquad\left.  + \lambda(d_{ijk} b^i b^j
  b^k -6)\right] \\ \label{resultgauge} S_4^{\textnormal{gauge}} &=& -
\frac{1}{16\pi\alpha_{\textnormal{GUT}}} \int
d^4x\sqrt{-g_4}\,\left[e^\phi\left(\tr F_{(0)}^2+\tr F_{(N+1)}^2
  \right) -\frac{1}{2}\sigma\epsilon_{\mu \nu \rho \gamma}\left(F^{\mu
      \nu}_{(0)} F^{\rho \gamma}_{(0)} + F^{\mu\nu}_{(N+1)}F^{\rho
      \gamma}_{(N+1)}\right) \right. \\ \nonumber
&&\left.\qquad\qquad\qquad\qquad\qquad +\sum_{p=1}^{N}\left(
    [\textnormal{Im}{\Pi}]_{(p)uw} E^u_{(p)}E^w_{(p)} -
    \frac{1}{2}[\textnormal{Re}{\Pi}]_{(p)uw} \epsilon_{\mu \nu \rho
      \gamma} E^{u\mu \nu}_{(p)}E^{w\rho\gamma}_{(p)}\right) \right]
\\ \label{resultmatter} S_4^{\textnormal{matter}} &=& -\int
d^4x\sqrt{-g_4}\,\sum_{p=0,N+1} \left[\frac{1}{2}\left( e^{-\beta}
    G_{(p)MN} DC_{(p)}^{Mx} D \bar{C}_{(p)x}^{N} - 2 e^{-2 \beta}
    G_{kl}\omega^{(p)k}_{1\mu} \partial^{\mu} \chi^l
  \right. \right. \\ \nonumber &&
\left. \left.\qquad\qquad\qquad\qquad\qquad + e^{-\phi -2
      \beta}G_{(p)}^{MN} \frac{\partial W_{(p)}}{\partial
      C^{Mx}_{(p)}} \frac{\partial \bar{W}_{(p)}}{\partial
      \bar{C}^M_{(p)x}} + e^{-2 \beta} {\rm tr}(D^2_{(p)})\right)
\right] \eea Here, $\kappa_{4}$ is the four-dimensional Planck
constant which is related to its 11-dimensional and 5-dimensional
counterparts by $\kappa_{4}^{2}=\kappa_{11}^{2}/\pi\rho
v=\kappa_{5}^{2}/\pi\rho$, where $v$ is the Calabi-Yau reference
volume and $\pi \rho$ is the interval length.  We remind the reader
that, in terms of quantities appearing elsewhere in this paper,
$z_{(p)} = y_{(p)}/ \pi \rho$, $\nu_{(p)} = s_{(p)}/ \pi \rho$, $\phi
= \ln V_0$, $\beta = \ln b_0$ and $\sigma$ is the dual of $H_{\mu \nu
  \rho}$.  The one-forms $\omega^{(p)k}_{1\mu}$ in the matter field
part of the above action are the Chern-Simons forms associated to the
currents~\eqref{J2}, that is, $d\omega^{(p)k}_1=J^{(p)k}_2$.  Finally,
the functions $W_{(p)}$ are given by \eqref{W} in Appendix B.
The second action in \eqref{z1} is much simpler and found to be
\begin{equation}
  S_{\delta^{1}}^{(1)} = -\frac{1}{2
    \kappa^2_4} \int d^4x\sqrt{-g_4}\, \left[ \frac{2\epsilon_0}{(\pi \rho)^2}e^{-\phi -2\beta}
    b^k\delta_k\right]
\label{z3}
\end{equation}
Recall that the quantities $\delta_k$, defined in \eqref{delta},
represent the differences of the tensions and charges of the anti three-brane. 
The complete action \eqref{z1}, in fact, remains valid if one replaces an arbitrary
number of branes with anti-branes. Then $\delta_k$ should
be interpreted as the sum of all anti-brane tensions minus the sum
of all anti-brane charges all divided by two.

Significant results can be gleaned merely by 
inspecting this action.

\begin{itemize}
\item First notice how close the action is, at this order, to the
  ${\cal N}=1$ supersymmetric result. Recall that formally switching
  off the supersymmetry-breaking effect of the anti-brane is achieved
  by setting $\delta_k\rightarrow 0$.  It follows that the
  $S_{\delta^{0}}^{(0,1)}$ portion of the action is identical to the bosonic
  part of the usual ${\cal N}=1$ supersymmetric
  theory~\cite{Lukas:1997fg,Lukas:1998hk,Derendinger:2000gy,Brandle:2001ts}. In
  particular, the kinetic term of the anti-brane's position modulus,
  which appears in $S_{\delta^{0}}^{(0,1)}$, is identical to that of a brane
  of the same tension. Indeed, if we define the scalar components of
  superfields in the standard way~\cite{Brandle:2001ts} by \bea S&=&
  e^\phi+\epsilon_0e^\beta\sum_{p=1}^{N}\,(\tau^{(p)}_{k}b^{k})z_{(p)}^2
  +i \left(\sigma+2\epsilon_0\sum_{p=1}^{N} \tau_{k}^{(p)} \chi^{k}
    z_{(p)}^2 \right) \label{Sdef}\\
  &=&e^\phi+i\sigma +\epsilon_0\sum_{p=1}^{N} \tau_{k}^{(p)}z_{(p)}^2T^{k} \\
  \label{Tdef}
  T^{k} &=& e^\beta b^{k}+2i\chi^{k} \\
  Z_{(p)} &=& \tau_{k}^{(p)}b^{k}e^\beta
  z_{(p)}+2i\tau_{k}^{(p)}(-n^{k}_{(p)}\nu_{(p)}
  +\chi^{k} z_{(p)}) \\
  &=&z_{(p)} \tau^{(p)}_{k}T^{k}-2i\tau_{k}^{(p)}n^{k}_{(p)}\nu_{(p)},
  \,\label{Zdef} \eea then  $S_{\delta^{0}}^{(0,1)}$ is reproduced as
  the bosonic part of the  ${\cal N}=1$ supersymmetric theory with K\"ahler potential
\begin{equation}\label{scalKaehlerPot}
   \k_{4}^{2}\, K_{\rm scalar}=K_{D}+K_{T}+\cK+K_{\rm matter}\; ,
\end{equation}
where
\begin{eqnarray}\label{K}
  K_{D}&=&-{\rm ln}\left[S+\bar{S}-\epsilon_0\sum_{p=1}^{N}
          \frac{(Z_{(p)}+\bar{Z}_{(p)})^{2}}{\tau_{k}^{(p)}(T^{k}+\bar{T}^{k})}
          \right]\,,\\
  K_{T}&=&-{\rm ln}\left[\frac{1}{48}d_{klm}(T^{k}+\bar{T}^{k})(T^{l}+\bar{T}^{l})
                                        (T^{m}+\bar{T}^{m})\right] \\
  \cK({\mathfrak z})&=&- \mbox{ln} \left[2 i ({\mathcal G}-\bar{{\mathcal G}})
    - i ( \mathfrak{z}^{p}-\bar{\mathfrak{z}}^{p})\left(\frac{\pt
    \mathcal{G}}{\pt\mathfrak{z}^{p}}+\frac{\pt
    \bar{\mathcal{G}}}{\pt\bar{\mathfrak{z}}^{p}}\right) \right]\\
 K_{\rm matter}&=&e^{K_{T}/3}\sum_{p=0,N+1} G_{(p)MN}
   C^{Mx}_{(p)}\bar{C}_{(p)x}^{N} \; ,
\end{eqnarray}
with superpotential for the matter chiral multiplets given by
\begin{equation}
 W_{(p)}=\sqrt{4\pi\alpha_{\rm GUT}}\sum_{I,J,K}\lambda_{IJK}f_{xyz}^{(IJK)}
        C_{(p)}^{Ix}C_{(p)}^{Jy}C_{(p)}^{Kz} 
\label{z5}
\end{equation}
and with gauge kinetic functions
\begin{eqnarray}
 f_{(p)}&=&S\; ,\quad p=0,N+1\\
 f_{(p)uv}&=&\Pi_{(p)uv}\; ,\quad p=1,\dots ,N \label{fdef}\; .
\end{eqnarray}
This is exactly the standard result for heterotic M-theory without
anti-branes~\cite{Brandle:2001ts}. Note that the superpotential
$W_{(p)}$ leads to a non-vanishing potential energy term for the
matter scalars $C_{(p)}^{Ix}$. However, the matter independent
potential energy of the dilaton and moduli fields vanishes in the
formal limit $\delta_k \rightarrow 0$, as it must.

\item Since $S_{\delta^{1}}^{(1)}$ in \eqref{z3} is proportional to
  $\delta_{k}$, we expect that this part of the action breaks the
  ${\cal N}=1$ supersymmetry.  To this order,
  $\kappa_{11}^{\frac{2}{3}}$, the supersymmetry breaking part of the
  four-dimensional bosonic effective action is very simple, merely
  adding the single term
\begin{equation}
  {\cal V}_1=\kappa_4^{-2}\frac{\epsilon_0}{(\pi \rho)^2}e^{-\phi -2\beta}b^k\delta_k\; \label{V1}
\end{equation}
to the potential energy. Note that this breaking term is a direct
consequence of the presence of the anti-brane. Even though
${\cal{V}}_{1}$ is not supersymmetric, it can still be expressed in
terms if the scalar fields $S$ and $T^{k}$ defined in \eqref{Sdef} and
\eqref{Tdef} respectively. We find that
\begin{equation} {\cal{V}}_{1}= \kappa_4^{-2}\frac{\epsilon_0}{(\pi
    \rho)^2} (T^k+\bar{T}^k) \delta_k e^{K_T+K_D} \;.
\label{c2}
\end{equation}
This potential term corresponds to twice the tension of the
anti-brane, precisely what one would add as a ``raising term'' in a naive
probe brane analysis.  Hence, our result lends additional weight to the
probe brane approach which, for example, is frequently used within the
context of IIB models.  However, we will see in the next section that
the probe brane approach breaks down completely at order
$\kappa_{11}^{\frac{4}{3}}$, where new contributions to the potential
appear. In the case where $h^{1,1}(X)=1$, expression \eqref{c2}
simplifies to
\begin{equation} {\cal{V}}_{1}= 8 \kappa_4^{-2}\frac{\epsilon_0}{(\pi
    \rho)^2} \delta \frac{1}{(S+\bar{S}) (T+\bar{T})^2} + O(\epsilon_0^2) \;.
\label{c3}
\end{equation}
Note that this is exactly the anti-brane induced potential computed
in~\cite{Ovrut:20061} and used to demonstrate the possibility of a
meta-stable dS vacuum with a small cosmological constant within the
context of an MSSM heterotic
standard model.\\

To summarize, including the full back reaction of the anti-brane to
first order in $\kappa_{11}^{\frac{2}{3}}$, the bosonic effective
theory~\eqref{resultmoduli}--\eqref{z3} is simply the
bosonic part of the supersymmetric theory as specified
by~\eqref{Sdef}-\eqref{fdef} above plus the single potential
contribution~\eqref{V1}.  We should also stress that, while the
bosonic part of the effective action can be interpreted as a
supersymmetric theory plus a raising term, things are not quite so
simple for the fermionic terms.  For example, the fermionic partner of
the anti-brane modulus $\bar{z}$ has the opposite chirality from the
fermions which originate from the boundaries and branes.  The above
supersymmetric theory would, therefore, not correctly reproduce terms
in the effective action which involve anti-brane fermions.

\item The ``raising potential''~\eqref{V1} is proportional to
  $e^{\frac{\phi}{3}} b^k\delta_k$ which, up to a constant, is the
  volume of the two-cycle within the Calabi-Yau space wrapped by the
  anti five-brane\footnote{To see this, note using~\eqref{delta},~\eqref{last2}
  and the relation $b^k=V^{-1/3}a^k$ discussed in Appendix A that $e^{\phi/3}
  b^k\delta_k=-\bar{\beta}_{k}a^k$. It then follows from~\eqref{last1} 
  and~\eqref{last3} that $e^{\phi/3}
  b^k\delta_k=-\int_{C^{(\bar{p})}}{\omega}$, which 
  is the volume of the two-cycle
  $C^{(\bar{p})}$ on which the anti five-brane is wrapped}.  
  For a Calabi-Yau space with sufficiently many
  K\"ahler moduli (two may be sufficient) we can shrink this cycle to
  zero size while keeping the overall Calabi-Yau volume $e^\phi$ (as
  well as the orbifold size $e^\beta$) large. In this limit, the
  anti-brane potential and the supersymmetry breaking it induces
  become small. As we will see, this is no longer true once order
  $\kappa_{11}^{\frac{4}{3}}$ corrections to the anti-brane potential
  are included.
      
\item The next thing to note about the action
  \eqref{resultmoduli}--\eqref{z3} is that it contains no
  potential terms involving the anti three-brane or three-brane
  position moduli. This may be surprising as our vacuum is no longer a
  BPS state and one would not expect the gravitational and form charge
  interactions of our extended objects to cancel each other
  out. Indeed, we will see that for terms of order
  $\kappa_{11}^{\frac{4}{3}}$ there is a non-zero force between these
  branes. It is easy to see why these terms do not arise at order
  $\kappa_{11}^{\frac{2}{3}}$. Physically, we expect a
  ``Coulomb-type'' force between, say, the anti three-brane and a
  three-brane.  This force would be proportional to the charge of {\it
    both} branes, as usual. Such terms, which are second order in the
  brane charges, are always higher order in the
  $\kappa_{11}^{\frac{2}{3}}$ expansion as well and, so, do not appear
  here. Another way of saying this is to point out that one
  contribution to such a force would be obtained by substituting the
  warping in the bulk fields caused by one brane into the world-volume
  action of the other. The warping is first order in
  $\kappa_{11}^{\frac{2}{3}}$ as are the brane world volume theories
  and, hence, this gives rise to a second order result.

  This point has important implications for heterotic moduli
  stabilization in the presence of anti five-branes as will be 
  discussed in the next section.
\end{itemize}

\subsection{A Duality Amongst the Effective Theories}

In the action presented in the previous subsection, different four
dimensional theories are obtained by making different choices for the
set of integers ${\tau^{(p)}_k}$. One particularly useful duality
amongst these theories, from the point of view of later sections of
this paper, is that which is the lower dimensional manifestation of
the symmetry between the orbifold fixed planes. In other words if, in
five dimensions, we were to view Fig. \ref{fig1} from the other side
of the page then the two orbifold fixed planes and the left to right
ordering of the branes would be swapped. Such a trivial change of
viewpoint can clearly not change the physics of the situation and so
should be represented as symmetries or dualities swapping around these
objects in the various descriptions of the system in different
dimensionalities.

In terms of the component fields in our four dimensional action the
relevant transformations are as follows.\bea \label{ztrans} z_{(p)}
&\rightarrow& 1 - z_{(p)} \\ \label{nutrans} \nu_{(p)} &\rightarrow& -
\nu_{(p)} \\ \label{sigmatrans} \sigma &\rightarrow& \sigma +2
\epsilon_0 \sum_{p=1}^N 2 \tau_k^{(p)} n^k_{(p)} \nu_{(p)}
\\ \label{indextrans} (p) &\rightarrow& (N+1-p) \eea All other
quantities are invariant.  The transformations \eqref{ztrans} and
\eqref{indextrans} are transparent in their physical content - they
correspond in a simple manner to inverting the diagram found in
Fig. \ref{fig1}. The transformations of the axions, \eqref{nutrans}
and \eqref{sigmatrans}, are then the changes that are required in the
definitions of the remaining four dimensional fields in order to make
the duality manifest.

In terms of the `superfields' defined in \eqref{Sdef}-\eqref{Zdef} the
transformations \eqref{ztrans}-\eqref{indextrans} become,
\bea \label{Ztranssuper}
Z_{(p)} &\rightarrow& \tau_k^{(p)} T^k - Z_{(p)} \;,\\
S &\rightarrow& S + \epsilon_0 \sum_{p=1}^N \tau_k^{(p)} T^k - 2
\epsilon_0 \sum_{p=1}^N Z_{(p)}\;,\\ \label{indextranssuper} (p)
&\rightarrow& (N+1-p) \;, \eea with all other quantities being
invariant.

The results presented in the previous subsection, and indeed in the
rest of this paper, are invariant under these transformations. This
provides one of the checks which we apply to our results.

In general, the above transformations constitute a set of dualities
among the possible four dimensional theories rather than a symmetry of
a given theory. This is due to the non-trivial action implied by
\eqref{indextrans} on the parameters $\tau^{(p)}_k$.

\section{Some Results at Second Order}
\label{secondorder}

Calculating terms at second order in the $\kappa_{11}^{\frac{2}{3}}$
expansion in heterotic M-theory is, a priori, a difficult thing to
do. The reason for this is that the full 11-dimensional theory is not
known at order $\kappa^{4/3}$ and so, in general, one cannot calculate
such terms in the effective theory.

However, it has been shown~\cite{Lukas:1997fg} that this argument is
too naive and that there are {\it some \rm} quantities which can be
reliably calculated at second order.  Examples are the second order
corrections to the matter Lagrangian and the threshold corrections to
the gauge kinetic functions \cite{Lukas:1997fg}. These terms are
calculated by showing that none of the unknown quantities in the
11-dimensional theory can possibly contribute to the relevant pieces
of the four-dimensional effective theory.  In this section, we show
that similar arguments can be employed in our case to calculate the
second-order terms in the anti-brane potential and the gauge kinetic
functions.

\subsection{Second Order Potential Terms}

In subsection \ref{01order}, we showed that the contributions to the
potential energy describing the perturbative forces between the branes
and anti-branes are at least second order in
$\kappa_{11}^{\frac{2}{3}}$. Furthermore, we argued that these forces
are necessarily non-vanishing at this order. This could be very
problematic since an understanding of these forces is necessary for a
full discussion of moduli stabilization. Fortunately, it turns out
that the relevant potential terms are exactly of the form which can be
reliably calculated at second order, although for somewhat different
reasons than more conventional examples.

Let us describe in detail how the calculation of the second order
terms in the potential can be accomplished. The crucial point is that
the brane positions $z_{(p)}$ are embedding coordinates.  Hence, there
are no terms at any order in the five-dimensional action which
explicitly involve $z_{(p)}$. Rather, the $z_{(p)}$ dependence in the
four-dimensional brane actions comes from only two sources. These are
1) the $z_{(p)}$ dependence of the induced metric
\eqref{inducedmetric} and, in general, any pulled back quantity, and
2) the background warping, which depends on the position of the branes
and anti-branes.  The first of these, the induced metric and
pull-backs, all involve four-dimensional derivatives. Hence, such
$z_{(p)}$ dependence cannot lead to potential energy terms in the
four-dimensional effective theory.  Therefore, the only possible
source of potential energy terms involving $z_{(p)}$ are those
obtained by substituting a $z_{(p)}$ dependent piece of the background
into a term in the five-dimensional action which does not contain
four-dimensional derivatives. Now, $z_{(p)}$ only appears in the
warping, that is, it appears at first order in
$\kappa_{11}^{\frac{2}{3}}$ in the background solution. Substituting
this into the brane actions, which are already of order
$\kappa_{11}^{\frac{2}{3}}$, leads to second order terms. Hence,
unknown correction terms to the brane action at order
$\kappa_{11}^{4/3}$ or higher would lead to terms in the potential
energy of order three or higher, which we do not consider
here. Similar comments can be made about unknown terms in the bulk
action. Thus, we cannot obtain a second order contribution to the
$z_{(p)}$ dependent potential by substituting the background warping
into the unknown second order five-dimensional action terms. In fact,
simple dimensional analysis in the 11-dimensional theory shows that no
bosonic order $\kappa^{4/3}$ terms on the branes and boundaries can be
written down at all, while the relevant bulk terms are
known. Therefore, even the $z_{(p)}$ independent parts of the second
order potential energy can be reliably computed. In principle, one
expects $z_{(p)}$ dependence in the (unknown) second order warping as
well.  However, an examination of the five-dimensional bulk action
reveals only two types of terms which do not contain four-dimensional
derivatives. The first of these are terms which do not involve derivatives
at all. However, these are all at least first order and, as such,
would give rise to third and higher order terms if a second order
background were to be substituted into them. The second type of term
is quadratic in single $y$ derivatives acting on the
background\footnote{Although the gravitational action naively involves
  double $y$ derivatives, these are removed upon integration by parts
  and a careful consideration of the Gibbons-Hawking boundary
  terms.}. If we act on the second order background with one of these
$y$ derivatives then, to obtain a second order term, we would require
that the other $y$ derivative acted upon the zeroth order
background. But the zeroth order background is $y$ independent and,
hence, such terms also do not contribute to the four-dimensional
effective theory.

Thus, we find that the only possible sources of second order potential
energy terms in the four-dimensional effective action are the following.
\begin{itemize}
\item The  first order background substituted twice into the 
 zeroth order bulk action.
\item The first order background substituted into the first order world volume
  action of extended sources.
\item The $\hat{\beta}^2/V^2$ term in the five-dimensional bulk action. 
\end{itemize}
The crucial point is that we know all of the relevant quantities
required to calculate these contributions.  We may proceed, then, to
calculate the ${\cal O}(\kappa_{11}^{\frac{4}{3}})$ contribution to
the potential energy. The calculation does not involve any further
subtleties than those already mentioned. Therefore, we simply state
the result. Terms of this order contribute $S_{\delta^{1,2}}^{(2)}$ to
the total action. The subscript and superscript indicate that this
contains terms of both first and second power in $\delta_{k}$ and
second order in $\kappa_{11}^{\frac{2}{3}}$ respectively. Part of the
Lagrangian density of $S_{\delta^{1,2}}^{(2)}$ is a potential energy
${\cal{V}}_{2}$, which we find to be \bea
\label{VY}
{\cal V}_2 &=& \kappa^{-2}_4 \frac{\epsilon_0^2}{(\pi\rho )^2}
e^{-\beta -2 \phi}\, G^{kl}\delta_l\left[
  \sum_{p=0}^{\bar{p}-1}\tau^{(p)}_k\bar{z}
  -\sum_{p=\bar{p}+1}^{N+1}\tau^{(p)}_k\bar{z}
  -\sum_{p=0}^{\bar{p}-1}\tau^{(p)}_kz_{(p)}\right.\nn\\
&&\qquad\qquad\qquad\qquad\qquad\left.
  +\sum_{p=\bar{p}+1}^{N+1}\tau^{(p)}_kz_{(p)} +
  \sum_{p=0}^{N+1}\tau^{(p)}_k(1-z_{(p)})z_{(p)}-\frac{2}{3}\delta_k\right].
\eea Recall that $p=\bar{p}$ labels the anti three-brane and $\bar{z}$
is the normalized anti-brane position modulus. Equivalently, in terms
of the fields defined in~\eqref{Sdef}-\eqref{Zdef}, this second order
term can be written as follows.
\bea \label{add3} {\cal{V}}_{2}= \kappa_4^{-2} \frac{\epsilon_0^2}{(\pi \rho)^2}
e^{K_T +2 K_D} K_T^{\bar{k} l} \delta_l \left[ \sum_{p=0}^{\bar{p}-1}
  \tau^{(p)}_k \frac{Z_{(\bar{p})} + \bar{Z}_{(\bar{p})}}{\bar{\tau}_m
    (T^m +\bar{T}^m)} - \sum_{p=\bar{p}+1}^{N+1} \tau^{(p)}_k
  \frac{Z_{(\bar{p})} + \bar{Z}_{(\bar{p})}}{\bar{\tau}_m (T^m
    +\bar{T}^m)} \right. \\ \left. \nonumber - \sum_{p=0}^{\bar{p}-1}
  \tau^{(p)}_k \frac{Z_{(p)} + \bar{Z}_{(p)}}{\tau_m^{(p)} (T^m
    +\bar{T}^m)} + \sum_{p=\bar{p}+1}^{N+1} \tau^{(p)}_k \frac{Z_{(p)}
    + \bar{Z}_{(p)}}{\tau_m^{(p)} (T^m +\bar{T}^m)} \right. \\
\left. \nonumber + \sum_{p=0}^{N+1} \tau^{(p)}_k \left( 1-
    \frac{Z_{(p)}+\bar{Z}_{(p)}}{\tau^{(p)}_m (T^m+\bar{T}^m)} \right)
  \frac{Z_{(p)}+\bar{Z}_{(p)}}{\tau^{(p)}_n (T^n+\bar{T}^n)}
  -\frac{2}{3} \delta_k \right].
\eea
In the above expression, $K_T^{\bar{k} l}$ is the inverse of 
$K_{T \bar{k}l} = \partial_{\bar{T}^k} \partial_{T^l} K_T$. 
The total potential of the theory, ${\cal V}$, is then given by
\begin{equation}
 {\cal V}={\cal V}_1+{\cal V}_2,
\end{equation}
where ${\cal V}_1$ and ${\cal V}_2$ are first and second order in
$\kappa_{11}^{\frac{2}{3}}$, respectively. The first order result,
${\cal V}_1$, has been presented in \eqref{V1} or, equivalently,
in~\eqref{c2}. Similarly, the second order result, ${\cal V}_2$, is 
given in component fields and superfields by \eqref{VY} and \eqref{add3} respectively.

As before, one can extract interesting physics simply by inspecting
these expressions.
\begin{itemize}
\item The physical interpretation of the first four terms in
  \eqref{VY} is clear. These two terms represent a force on
  the anti-brane.  This force is proportional to the anti-brane's
  charge and receives two contributions; first, a contribution
  proportional to the sum of the charges to the left of the anti-brane
  and second, a force in the opposite direction proportional to the
  sum of the charges to the right of the anti-brane. This simply
  represents the Coulomb attraction of the anti-brane to the
  brane-like charges, a force which is no longer zero in this non-BPS
  configuration. Similarly, the third and fourth terms are the forces
  the branes experience pulling them towards the anti-brane. Each of
  these terms is proportional to the charge of the brane of interest
  multiplied by the charge of the anti-brane. Note that in the first
  four terms in ${\cal V}_2$ the three-branes do not attract one
  another, that is, there are no terms quadratic in the brane (as
  opposed to anti-brane) charges.
\item The last two terms are more surprising since they are not of
  Coulomb type but have the same magnitude as the preceding ones.  As
  far as the authors are aware, the existence of such effects has not
  been previously discussed in the literature. These terms are a
  result of performing a full calculation involving back-reaction and
  would be difficult to guess from a probe calculation. They arise
  from the background warping. This appears in both the tensions of
  the branes and bulk terms, changing their energies.  In particular,
  recall that we have defined our zero mode quantities so that the
  orbifold average of the warping is zero. Since the warping depends
  upon the brane positions, this means we have to add to it a function
  of the $z_{(p)}$ such that the overall average is zero for any value
  of the brane coordinates. This ``normalization'' of the warping
  appears in the tensions terms equally for each extended
  object. Normally this does not result in a potential energy, since
  the tensions of the extended objects sum to zero. However, in the
  case where anti-branes are present the tension terms do not sum to
  zero. The brane positions will then try to adjust so as to minimize
  the warping normalization contribution to the potential energy. This
  is the source of the new potential term presented above. We
  emphasize at this point that this is {\it not} ambiguous in any
  way. Nor is it an artifact of our choices of normalization for the
  warpings. If we wished to change the normalization to remove this
  term, we would at the same time change terms in the first order
  action presented in the previous subsection, including the kinetic
  terms. This would, in fact, simply correspond to a field
  redefinition of the four-dimensional theory presented here. A field
  redefinition can not change the physical properties of a system. Our
  intuition as to the physics of these compactifications is built
  around our standard choices of bases in field space. We will
  maintain these choices in examining our results.  Clearly, including
  this potential energy term is vital when considering the
  stabilization of moduli in heterotic M-theory.
\item Another interesting point to make about the above potential is
  that it does not necessarily vanish as the size of the cycle on
  which the anti-brane is wrapped is taken to zero. Indeed, \eqref{VY}
  is not proportional to the volume $b^k\delta_k$ wrapped by the anti
  three-brane. It is easy to understand from the structure of the
  K\"ahler moduli indices why ${\cal V}_1$ had to be proportional to
  the anti-brane cycle, and why this is not necessarily the case for
  ${\cal V}_2$. Clearly, the full potential ${\cal V}$ must be
  proportional to the anti-brane charge $\delta_k$. The first order
  potential must be linear in the brane charges and, hence, linear in
  $\delta_k$. However, the K\"ahler index $k$ on $\delta_k$ must be
  contracted and the only objects available for this are the K\"ahler
  moduli $b^k$ \footnote{Naively one could also use the combination
    $n^k_{(p)} s_{(p)}$ and the $\chi^k$ from the axionic
    sector. However, these axionic fields all enjoy a shift symmetry
    and so can not appear in such potential terms.}. The second order
  potential, on the other hand, is bilinear in charges which allows
  for more complicated expressions. For example, bilinears can be
  contracted with the K\"ahler metric.  How can shrinking the cycle
  which the anti-brane wraps to a very small size not make the force
  between it and and the branes small? After all, the total charge of
  the object is controlled by the size of its world volume. The answer
  comes from an understanding of how the K\"ahler moduli $b^k$ are
  defined. They have been scaled so that, as we vary them, the overall
  volume of the manifold remains constant (this volume being
  controlled by $e^\phi$). In particular, this means that as we shrink
  a cycle down to a small size another must expand so as to keep the
  volume constant. If an M five-brane is wrapped on that cycle, then
  its total charge will increase. Since the force between the two
  objects is equal to the product of the charges, it is not entirely
  clear which effect will win in general. The above potential has to
  be analyzed for each case in order to answer this question.  The
  fact that the brane moduli dependent part of the potential is second
  order in $\epsilon_S$ means that, in any regime of moduli space
  where four-dimensional heterotic M-theory is a good description of
  the physical situation, this contribution to the potential is
  suppressed. As mentioned earlier, this is a good thing from the
  point of view of moduli stabilization. It means that it should be
  easier to balance this force on the brane and anti-brane positions
  against non-perturbative effects.
\item Note that the double derivative of this potential with respect
  to any of the brane position moduli is negative. That is to say
  that, depending on the configuration of charges under consideration,
  there may be a stationary point of the potential in the direction of
  the brane moduli. However, this stationary point is at best a saddle
  point. This property is illustrated in Fig.~\ref{fig3} for the case
  of a single anti-brane with no branes present.  Given the anomaly
  condition~\eqref{anomaly}, plus the fact that the anti-brane charge
  is negative, we have two qualitatively different situations: both
  boundary charges are positive, or one boundary charge is positive
  and the other negative. The potentials for both situations are
  plotted in Fig.~\ref{fig3}. 

  We should stress that this perturbative instability by no means
  implies that heterotic models with anti-branes are necessarily
  unstable. Given the relative weakness of the perturbative force,
  non-perturbative effects must be taken into account in a stability
  analysis. Since membrane instanton effects typically repel a brane
  from the boundaries \cite{otherpapers}, it is, in fact, likely that
  the anti-brane can be stabilized by a combination of perturbative
  and non-perturbative effects. This perturbative runaway to
  undesirable field values is no different from that of the dilaton in
  heterotic theory or the K\"ahler moduli in type IIB strings, and
  should be treated in the same manner.

\item Given the unexpected extra terms in the above result, it is
  important to do some non-trivial checks of the calculation. Indeed,
  many robust checks of this result are possible and have been
  performed. First of all, the standard results of heterotic M-theory
  should be reproduced if we set $\delta_k$ to zero. This is indeed
  the case (the potential vanishes). There are many symmetries which
  the potential must respect. For example the transformations
  \eqref{ztrans}--\eqref{indextrans} should leave the result
  invariant. This simply corresponds to the fact that what we call the
  left and right hand orbifold fixed points is a matter of
  convention. The above potential indeed exhibits this
  property. Furthermore, the result can also be explicitly
  re-calculated in different ways. For example, one can switch to a
  dual description of this system \cite{Brandle:2001ts} where the
  $\hat{\beta}$ terms in the bulk five-dimensional action are
  exchanged for a bulk five-form field strength. We have performed the
  calculation in this dual description and again reproduce the above
  potential.  In short, there are robust checks one can perform,
  despite the fact that the system is not supersymmetric. The above
  result for ${\cal V}$ passes all of these tests.
\end{itemize}
\begin{figure}\centering 
\includegraphics[height=7.5cm,width=10cm, angle=0]{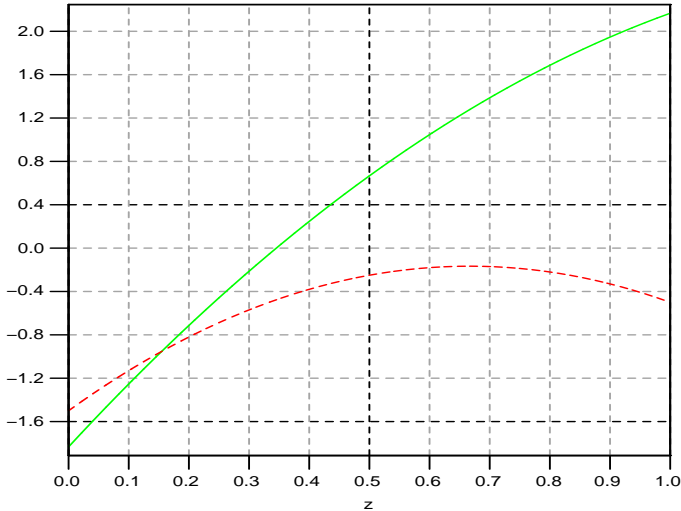}
\caption{Anti-brane potential in the absence of other branes (from the
  bracket on the RHS of \eqref{VY}), assuming $h^{1,1}(X)=1$. The
  solid, green curve corresponds to charges $(\beta^{(p)})=(3,-2,-1)$. In
  this case the anti-brane is attracted to the positively charged
  boundary at $z=0$. The red, dashed curve corresponds to charges
  $(\beta^{(p)})=(2,-3,1)$. The anti-brane is attracted to either one
  of the boundaries, depending on its position $z$.}
\label{fig3}
\end{figure}

\subsection{ Threshold Corrections to the Gauge Kinetic Functions}
\label{threshold}

Another set of quantities relevant for moduli stabilization which
arise at second order in $\kappa_{11}^{\frac{2}{3}}$ are the
threshold corrections to the gauge kinetic functions. Again, we are
fortunate. Standard arguments \cite{Lukas:1997fg} tell us that these are among
the few terms which can be reliably calculated at this order. The
arguments are slightly altered in the current situation, but are
essentially unchanged.

Consider the possible sources of second order contributions to the
gauge kinetic functions. One possible source is higher order terms
involving the gauge field strengths in the higher dimensional
action. Clearly, such terms can only occur on the boundaries where the
gauge fields are located. Dimensional analysis shows that such
contributions are either all higher order or, possibly, come with a
non-integer power of $\kappa_{11}^{2/3}$. This latter possibility is
probably forbidden by the supersymmetry of the five-dimensional theory
and, in any case, cannot mix with the terms we wish to
calculate. Therefore, we will assume that such terms do not
occur. 

Another possible source of second order terms in the gauge kinetic
functions are $\kappa_{11}^{\frac{4}{3}}$ terms in the
warping. Clearly, substituting such a piece of the background into the
boundary actions will give contributions that are higher order than we
are interested in. Substituted into the bulk theory, such second order
contributions to the warping will not contribute to the
four-dimensional effective action if there is a zero mode associated
with them. This is because the zero mode is defined to be the orbifold
average and, hence, the linear perturbation of order
$\kappa_{11}^{\frac{4}{3}}$ drops out upon performing the integration
over the orbifold direction. This is not quite true for some of the
background pieces in $G$ and ${\cal A}$. These could, in principle,
contribute to the kind of terms we are interested in. However, in the
supersymmetric case the results obtained were consistent with these
contributions being zero. In other words, the complex structure
determined from the moduli kinetic terms, and the requirement that the
gauge kinetic functions be holomorphic, are consistent with there
being no additional terms of this type. This is a highly non-trivial
statement and is unlikely to have occurred by chance. We will assume
that these terms are also absent in the non-supersymmetric case. This
assumption leads us to a result which is consistent with all of the
symmetries present in this situation, including being invariant under
the transformations \eqref{ztrans}--\eqref{indextrans}. This
constitutes a highly non-trivial check of its veracity.

The crucial point, once one has concluded that the above contributions
do not occur, is that all other ways of generating second order terms
in the gauge kinetic functions are explicitly calculable using pieces
of the action and the background solution that we know. One obvious
contribution arises by inserting the domain wall solution into the
boundary Yang-Mills actions. Other, more subtle, contributions arise
from the additional warping of bulk fields caused by fluctuations on
the boundary. For example, the four-dimensional metric receives a
warping that is bilinear in the boundary gauge field strengths. This
warping will lead to order an $\kappa_{11}^{\frac{4}{3}}$ correction
to the gauge kinetic function via cross terms with the domain wall
warping. The full expressions for the warpings of the bulk fields that
are needed in the calculation are given in
Appendix~\ref{App:Bianchi}. However, we find that all contributions
due to warping induced by field fluctuations drop out, just as they
did in the supersymmetric case~\cite{Lukas:1997fg}. This means that,
in the end, the threshold corrections to the gauge kinetic functions
are simply determined by the warping of the domain wall
solution. Explicitly, to order $\kappa_{11}^{\frac{4}{3}}$, we find
that the Yang-Mills part of the four-dimensional effective theory is
given by \bea \label{gkf2ndorder} S_4^{\textnormal{GKF}} &=&
\frac{-1}{32 \pi \alpha_{\textnormal{GUT}}} \int d^4 x \sqrt{-g_4}
\left[ \left( e^\phi + \epsilon_0e^\beta b^k \left( \sum_{p=0}^{N+1}
      \tau_k^{(p)} \left(z_{(p)}^2-2z_{(p)} \right) +
      \frac{4}{3}\delta_k \right) \right)
  \textnormal{tr} (F^2_{(0)}) \right. \nn\\
&& \left.\qquad\qquad\qquad\qquad\qquad + \left( e^\phi +
    \epsilon_0e^\beta b^k \left( \sum_{p=0}^{N+1}
      \tau^{(p)}_kz_{(p)}^2-\frac{2}{3}\delta_k\right)
  \right) \textnormal{tr} (F^2_{(N+1)})\right. \nn \\
&& \left.  - \frac{1}{2} \left( \sigma + 2\epsilon_0\left(
      \sum_{p=1}^{N} \beta^{(p)}_k \chi^k \left(z_{(p)}^2-2z_{(p)}
      \right) - \beta^{(N+1)}_k\chi^k + 2 \sum_{p=1}^{N}\tau^{(p)}_k
      (n^k_{(p)}\nu_{(p)}) \right)\right)\epsilon_{\mu \nu \rho
    \sigma} F_{(0)}^{\mu \nu} F_{(0)}^{\rho
    \sigma} \right. \nn\\
&& \left.  - \frac{1}{2} \left( \sigma +
    2\epsilon_0\left(\sum_{p=1}^{N}\beta^{(p)}_k\chi^kz_{(p)}^2 +
      \beta_k^{(N+1)} \chi^k\right)\right) \epsilon_{\mu \nu \rho
    \sigma} F_{(N+1)}^{\mu \nu} F_{(N+1)}^{\rho \sigma} \right].  \eea

Despite the fact that our theory is not supersymmetric, we can express
this result in terms of two `gauge kinetic functions'. One simply
defines the real and imaginary parts of these objects to be given by
the coefficients of the $\textnormal{tr} F^2$ and $\epsilon F F$ terms
as usual. We find that
\bea \label{f0comp} f_{(0)} &=& \left( e^\phi + \epsilon_0e^\beta b^k
  \left( \sum_{p=0}^{N+1} \tau_k^{(p)} \left(z_{(p)}^2-2z_{(p)}
    \right) + \frac{4}{3}\delta_k \right) \right) \\ \nonumber &&+ i
\left( \sigma + 2\epsilon_0\left( \sum_{p=1}^{N} \beta^{(p)}_k \chi^k
    \left(z_{(p)}^2-2z_{(p)} \right) - \beta^{(N+1)}_k\chi^k + 2
    \sum_{p=1}^{N}\tau^{(p)}_k (n^k_{(p)}\nu_{(p)}) \right)\right)
\\ \label{fN1comp} f_{(N+1)} &=& \left( e^\phi + \epsilon_0e^\beta b^k
  \left( \sum_{p=0}^{N+1}
    \tau^{(p)}_kz_{(p)}^2-\frac{2}{3}\delta_k\right) \right) +i \left(
  \sigma + 2\epsilon_0\left(\sum_{p=1}^{N}\beta^{(p)}_k\chi^kz_{(p)}^2
    + \beta_k^{(N+1)} \chi^k\right)\right) \eea
The explicit supersymmetry breaking in our theory
means that these functions will no longer be holomorphic 
when expressed in terms of the superfields defined in 
\eqref{Sdef}--\eqref{Zdef}. We will see this
explicitly below.

Some comments about this result are in order.
\begin{itemize}
\item As was the case for the first order results, this expression is
  valid when we change an arbitrary number of our branes into
  anti-branes. In this case, the quantities $\delta_k$ should be
  taken to be the sum of the anti-brane tensions.
\item The above result agrees with that of standard heterotic M-theory
  upon taking $\delta_k\rightarrow 0$. Despite the fact that these
  terms are second order in our expansion, they only contain a single
  power of the brane charges. This is necessarily the case, given that
  the gauge fields themselves only appear at order
  $\kappa_{11}^{\frac{2}{3}}$. This means that all of the threshold
  corrections to the gauge kinetic functions can be made small by
  tuning the $b^k$, $\chi^k$ and $\nu_{(p)}$ moduli appropriately. In
  fact, an inspection of \eqref{gkf2ndorder} reveals that it suffices
  to set the volume $\delta_kb^k$ of the anti-brane cycle and the
  associated linear combination $\delta_k\chi^k$ of the $\chi^k$
  axions to zero.

\item We can write the result as a supersymmetric piece plus explicit
  supersymmetry breaking terms, using the definition of superfields in
  \eqref{Sdef}--\eqref{Zdef}. This leads to a gauge kinetic function
  which is no longer holomorphic at second order in our
  expansions. \bea \label{gkfsuper1} f_{(0)} &=&S-\epsilon_0\left[
    \tau^{(N+1)}_{k}T^{k}+2\sum_{p=1}^{N}Z_{(p)} - \frac{2}{3}\delta_k
    (T^k + \bar{T}^k) \right. \\ && \;\;\;\; \quad\quad \quad \quad
  \quad \left. \nonumber + \delta_k (T^k-\bar{T}^k) \left(
      \left(\frac{Z_{(\bar{p})}+ \bar{Z}_{(\bar{p})}}{\bar{\tau}_k
          (T^k +\bar{T}^k)} \right)^2-2
      \frac{Z_{(\bar{p})}+\bar{Z}_{(\bar{p})}}{\bar{\tau}_k (T^k
        +\bar{T}^k)} \right) \right] \\ \label{gkfsuper2} f_{(N+1)}
  &=& S+ \epsilon_0\left[\tau_{k}^{(N+1)}T^{k}-\frac{1}{3}\delta_k
    (T^k+ \bar{T}^k) - \delta_k (T^k-\bar{T}^k)
    \left(\frac{Z_{(\bar{p})}+ \bar{Z}_{(\bar{p})}}{\bar{\tau}_k (T^k
        +\bar{T}^k)} \right)^2 \right] \eea The above result looks
  complicated, despite the fact that the component forms of the gauge
  kinetic functions, given in (\ref{f0comp},\ref{fN1comp}), are quite
  close to the usual result for heterotic M-theory. This is because,
  in many of the terms in the imaginary parts of these functions, it
  is the charge, and not the tension, of the extended objects which
  appears. The complex structure derived in subsection \ref{01order}
  contained the tensions rather than the charges. Therefore, we have
  to add contributions proportional to $\delta_k$ to the standard
  holomorphic quantity in order to correct this difference.
\item At first glance there appears to be a strange asymmetry in the
  results \eqref{gkfsuper1} and \eqref{gkfsuper2} between the $(0)$
  and $(N+1)$ fixed planes. For example, there is a linear term in
  $Z_{(p)}$ in \eqref{gkfsuper1} but not in \eqref{gkfsuper2}. This
  asymmetry is an artifact of the fact that we measure the position of
  the branes as a distance from the $(0)$ plane and the complicated
  way in which the component scalar fields are combined in the ${\cal
    N}=1$ structure \eqref{Sdef}-\eqref{Zdef}. In fact if we apply the
  transformations \eqref{Ztranssuper}-\eqref{indextranssuper},
  corresponding to inverting the $y$ direction in Fig.~\ref{fig1}, we
  find that the above gauge kinetic functions turn into one another as
  they should. This is a highly non-trivial test of our results.

  One could of course simply relabel the branes, swapping $(0)$ and
  $(N+1)$ if one so desired. Such a change of notation is useful in
  relating these results to some of the previous literature. We note
  in addition that either of the fixed planes can be the hidden
  sector. Which plays this role is determined by such factors as the
  choices of the tensions $\tau^{(p)}_k$.
\item Note that these corrections to the gauge kinetic
  functions arise at order $\epsilon_S$, while the brane forces
  appear at order $\epsilon_S^2$. Hence, if we wish to include the
  effects of gaugino condensation, this modification of the gauge
  kinetic functions is likely to be relevant in regions of moduli space
  where the exponential suppression is not too strong. This 
  appears to have been missed in the literature so
  far. This is not surprising, given the need for a full calculation of
  the back-reaction of the anti-brane before such corrections become
  evident.

  The importance of these terms has already been demonstrated in
  Fig.~\ref{fig2}, where we have plotted the warping of the Calabi-Yau
  volume across the orbifold. The values of this warping at the
  boundaries, $z=0,1$, correspond precisely to the real parts of the
  two gauge kinetic functions above. What we have seen is that the
  introduction of an anti-brane can turn the weakly coupled boundary
  into the strongly coupled one and vice versa.
  \end{itemize}


\section{A Simple Example}
\label{examplesection}

The reader interested in analyzing the physics of anti-branes in
four-dimensional heterotic M-theory may not wish to wade through the
details of the dimensional reduction presented in the proceeding
sections. With this in mind, we now provide a simple example
extracted from our general analysis.

Let us consider the case where there is only a single K\"ahler modulus
for the Calabi-Yau manifold, a single anti-brane and no branes in the
bulk.  Then action \eqref{z1} and the second order action
$S_{\delta^{1,2}}^{(2)}$ with potential \eqref{VY} become
\begin{equation}
S=S_{\delta^{0}}^{(0,1)}+S_{\delta^{1}}^{(1)}+S_{\delta^{1,2}}^{(2)},
\label{india1}
\end{equation}
where
\bea \label{egkt} S_{\delta^{0}}^{(0,1)}&=&-\frac{1}{2 \kappa_4^2}
\int d^4 x \sqrt{-g_4} \left[ \frac{1}{2} R_4 + \frac{3}{4} (\partial
  \beta)^2 + \frac{1}{4} (\partial \phi)^2 + \frac{1}{4} e^{-2 \phi}
  (\partial \sigma)^2 + 3 e^{-2\beta} (\partial \chi)^2 \right. \\
\nonumber &+& \left. 2 \epsilon_0 \bar{z} e^{-2 \phi} \partial
  \sigma \partial \bar{\nu}+ \frac{1}{2} \epsilon_0 \bar{\tau}
  e^{\beta - \phi} (\partial \bar{z})^2 + 2 \epsilon_0 \bar{\tau}
  e^{-\phi- \beta} \left( \chi^2 (\partial \bar{z})^2 - 2
    \frac{\chi}{\bar{\tau}} \partial \bar{\nu} \partial \bar{z} +
    \frac{1}{\bar{\tau}^2} (\partial \bar{\nu})^2 \right) \right]
\label{india2}
\\
S_{\delta^{1}}^{(1)}&=&-\frac{1}{2 \kappa_4^2} \int d^4 x \sqrt{-g_4}
\left[\frac{2 \epsilon_0}{(\pi \rho)^2} \delta e^{-2 \beta -
    \phi}\right]
\label{india3}
\eea
and
\bea
  S_{\delta^{1,2}}^{(2)}= -\frac{1}{2 \kappa_4^2} \int d^4 x
  \sqrt{-g_4} \left[ \frac{2 \epsilon_0^2}{3 (\pi \rho)^2}
    e^{-\beta -2 \phi}\delta \left(
      (\tau_{(0)}-\tau_{(2)}) \bar{z} + \bar{\tau} (1-
      \bar{z})\bar{z} + \tau_{(2)} - \frac{2}{3} \delta
    \right)\right].
\label{india4}
\eea
Here we have only included the potential terms in
$S^{(2)}_{\delta^{1,2}}$, as described earlier in the paper. This
action is valid when $\epsilon_0 e^{\beta - \phi} < 1$, where
$\epsilon_0 = (\pi \rho)^{\frac{4}{3}} 2 \pi v^{-\frac{1}{3}} \left(
  \frac{\kappa_4}{4 \pi}\right)^{\frac{2}{3}}$ for the conventions
chosen here.  The system consists of four-dimensional gravity and six
scalar fields, $\phi$, $\beta$, and $\bar{z}$ as well as $\sigma$,
$\chi$ and $\bar{\nu}$.  The volume of the Calabi-Yau space in
fundamental units is given by $v e^{\phi}$, the separation of the
orbifold fixed planes in five dimensions by $\pi \rho e^{\beta}$ and,
finally, $\bar{z}$ is the position modulus of the anti-brane. In these
expressions $v$ and $\pi \rho$ are reference constants, which can be chosen 
arbitrarily. They define the physical meaning of the moduli $\phi$ and
$\beta$. The remaining three scalars are the axions corresponding to
these fields. The fields $\sigma$ and $\chi$ descend from bulk form
fields in higher dimensions and $\bar{\nu}$ from a field living on the
worldvolume of the anti-brane. A collision of the anti-brane 
with one of the two fixed planes occurs
when the anti-brane modulus takes the value $\bar{z}=0$ or
$\bar{z}=1$. The only quantities appearing in the action which remain
to be explained are then $\bar{\tau}$, $\tau_0$ and $\tau_2$. These
are integers which determine the tensions of the anti-brane and the
orbifold fixed planes which the anti-brane collides with at
$\bar{z}=0$ and $\bar{z}=1$ respectively. There are two restrictions
on the choices which can be made for these parameters. First,
$\bar{\tau}$ must be positive. Second, the three tensions must obey
the condition $\tau_0 - \bar{\tau} + \tau_2 =0$.

Action \eqref{india1} can also be expressed in terms of the
`superfields' defined in \eqref{Sdef}-\eqref{Zdef}. For this simple
example, these become
\bea \label{india5a} S&=& e^\phi+ \epsilon_0 \bar{\tau}
e^\beta\bar{z}^2 +i \left(\sigma+2 \epsilon_0 \bar{\tau} \chi
  \bar{z}^2 \right) \\
T &=& e^\beta +2i\chi \\
Z &=& \bar{\tau} e^\beta \bar{z}+2i (\bar{\nu} +\chi \bar{\tau}
\bar{z}). 
\label{india5c}
\eea
First consider $S_{\delta^{0}}^{(0,1)}$ in \eqref{egkt}. This part
of the action is independent of the parameter $\delta$. Hence, it is
supersymmetric and can be expressed in terms of a
K\"ahler potential and a superpotential. Since this simplified theory
only contains the dilaton and moduli, the superpotential vanishes. We
find the K\"ahler potential to be
\begin{equation}
\kappa_4^2 K = -3 \log
\left[T+\bar{T}\right] - \log \left[S+\bar{S} - \epsilon_0
  \frac{(Z+\bar{Z})^2}{\bar{\tau} (T+\bar{T})} \right] + \log 8.
\label{india6}
\end{equation}
In calculating the kinetic terms, the results obtained from this
expression are valid to first order in $\epsilon_0 e^{\beta-\phi}$.

Now consider $S_{\delta^{1}}^{(1)}$ and $S_{\delta^{1,2}}^{(2)}$.
These terms are proportional to at least one power of $\delta$. Hence,
they are explicitly associated with the non-supersymmetric part of the
theory and we do not write them in terms of a K\"ahler potential and a
superpotential. Be that as it may, they can be expressed in terms of
the scalar fields in (\ref{india5a}-\ref{india5c}). Recognizing that
\eqref{india3} and \eqref{india4} contribute potential energy terms
${\cal{V}}_{1}$ and ${\cal{V}}_{2}$ respectively, we find that the
total potential induced by the anti five-brane is
\begin{equation}
{\cal V} = {\cal V}_1+{\cal V}_2,
\label{india7}
\end{equation}
where
\bea {\cal V}_1 &=& \kappa_4^{-2} \frac{\epsilon_0}{(\pi\rho)^{2}}
\delta (T+\bar{T}) e^{\kappa_4^2 K} \\ {\cal V}_{2} &=& \frac{1}{24}
\kappa_4^{-2} \frac{\epsilon_0^2}{(\pi\rho)^{2}} \delta e^{2
  \kappa_4^2 K}(T+\bar{T})^5 \left[ 2 \frac{\tau_{(0)}}{\bar{\tau}}
  \frac{Z+\bar{Z}}{T+\bar{T}} - \frac{1}{\bar{\tau}}
  \left(\frac{Z+\bar{Z}}{ T+\bar{T}}\right)^2 +\tau_{(2)} -\frac{2}{3}
  \delta \right]. \eea
This simple example includes the essentials of the features of the
potential outlined in the main text. In particular, while the first
term of \eqref{india4} represents the Coulomb forces acting on the
anti-brane, the second, third and fourth terms are the new
contributions to the potential that we have discussed. 

The gauge kinetic functions in this simple example are the
following. \bea f_{(0)} &=& e^{\phi} + \epsilon_0 e^{\beta} \left(
  \bar{\tau} (\bar{z}^2- 2\bar{z}) + \frac{4}{3} \delta \right) + i
\left( \sigma + 2\epsilon_0 (- \bar{\tau} \chi (\bar{z}^2 -2 \bar{z})
  - \tau_{(2)}
  \chi + 2 \bar{\nu} ) \right)  \; \\
f_{(2)} &=& e^{\phi} + \epsilon_0 e^{\beta} \left( \bar{\tau}
  \bar{z}^2 - \frac{2}{3} \delta \right) + i \left(\sigma + 2
  \epsilon_0 \left( -\bar{\tau} \chi \bar{z}^2 + \tau_{(2)} \chi
  \right) \right). \;\eea These expressions correspond to the orbifold
fixed planes at $\bar{z}=0$ and $\bar{z}=1$ respectively. The real
parts of these functions reproduce the inverse square of the gauge
couplings of the gauge fields living upon the fixed planes, while the
imaginary parts reproduce the theta terms. Again, one can rewrite
these results in terms of the field definitions
\eqref{india5a}-\eqref{india5c}. The result is \bea \label{eggkf0}
f_{(0)} &=& S - \epsilon_0 \left[ \tau_{(2)} T +2 Z - \frac{2}{3}
  \delta (T+\bar{T}) + \delta (T-\bar{T}) \left( \left(
      \frac{Z+\bar{Z}}{\bar{\tau} (T+\bar{T})} \right)^2 - 2
    \frac{Z+\bar{Z}}{\bar{\tau} (T+\bar{T})}\right) \right]
\\ \label{eggkf2} f_{(2)} &=& S + \epsilon_0 \left[ \tau_{(2)} T
  -\frac{1}{3} \delta ( T+\bar{T}) - \delta (T-\bar{T}) \left(
    \frac{Z+\bar{Z}}{\bar{\tau} (T+\bar{T})}\right)^2\right]. \eea
Note that the gauge kinetic functions are holomorphic in these complex
fields in the formal limit that $\delta \rightarrow 0$. This is as it
should be, since this corresponds to turning off the supersymmetry
breaking induced by the anti-branes. The terms proportional to
$\delta$ break supersymmetry and, accordingly, are not holomorphic in
the fields \eqref{india5a}-\eqref{india5c}. The fact that no single
choice of complex fields is possible which simultaneously allows the
kinetic terms in \eqref{egkt} to be reproduced from a K\"ahler
potential and makes the functions \eqref{eggkf0} and \eqref{eggkf2}
holomorphic demonstrates that the presence of the anti-brane breaks
supersymmetry explicitly. The modifications to the gauge kinetic
functions due to the presence of the anti-brane are crucial in any discussion of gaugino
condensation. These corrections can even exchange the weakly and
strongly coupled fixed planes, interchanging what would be naively
thought of as the visible and hidden sectors. Note that, for suitable
choices of the parameters described above, either fixed plane can be
the hidden sector.

Action \eqref{india1}, together with the gauge kinetic functions
\eqref{eggkf0} and \eqref{eggkf2}, can be used as the starting point for
analyzing the physics of anti-branes in perturbative heterotic
M-theory.

\section{Conclusions}
\label{Conclusion}

In this paper, we derived the bosonic four-dimensional effective
theory for heterotic M-theory in the presence of M five-branes and
{\it anti M five-branes}. The starting point of our analysis is the
five-dimensional action of heterotic M-theory, where
the M five-branes appear as three-branes. We have explicitly computed the case
with an arbitrary number of three-branes but only one anti three-brane. However, it is 
straightforward to generalize our results to an arbitrary number of
anti three-branes. 

We first found a suitable background solution to
five-dimensional heterotic M-theory on which to reduce to four-dimensions.
This solution is a non-supersymmetric domain wall, presented
in Section~\ref{dw}, which is a generalization of the BPS domain wall
background in the supersymmetric case, that is, in the absence
of anti-branes. We found that this domain wall solution can be computed
as an expansion in the strong coupling parameter $\epsilon_S$, and we
presented the result up to first order in this parameter.
A new feature induced by the anti-brane is warping
quadratic in the orbifold coordinate. This occurs in addition to the linear warping
which is characteristic of the BPS domain wall in the supersymmetric theory.

We then computed the four-dimensional bosonic effective action on this
domain wall background as an expansion in
$\kappa_{11}^{\frac{2}{3}}$. We stress that this calculation includes
the back-reaction effects of the branes as well as the anti-branes. In
particular, there is no assumption of a small anti-brane charge
underlying our calculation. To zeroth and first order in this
expansion, we found that the bosonic theory is given by the
supersymmetric result plus the addition of a single potential term.
This ``uplifting potential'' is generated by the anti-brane and has a
number of interesting properties. First of all, at this order, it is
independent of the brane position moduli and only depends on the
dilaton and the Calabi-Yau K\"ahler moduli. More precisely, it is
proportional to the volume of the cycle wrapped by the anti brane, as
well as to the strong coupling parameter $\epsilon_S$.  Since,
necessarily, $\epsilon_S\ll 1$ (for four dimensional heterotic
M-theory to be a good description of the system) the uplifting
potential is suppressed.

We have also calculated a number of terms in the four-dimensional
effective action at second order, that is, at order
$\kappa_{11}^{\frac{4}{3}}$. We have specifically focused on terms
where new qualitative features, not seen at lower order, occur.  In
particular, we explicitly calculated the second order contributions to
the anti-brane potential. This gives interactions between branes and
anti-branes. This potential is of order $\epsilon_S^2$ and, hence,
even further suppressed relative to the leading uplifting
potential. The $\epsilon_S$ suppression of the perturbative potential
has important implications for moduli stabilization, since the
stabilization mechanism involves an interplay between perturbative and
non-perturbative effects. For stabilization to occur, both types of
effects should be roughly comparable in size. This is greatly
facilitated by the suppression of the perturbative potential.  While
the second order brane potential includes the expected
``Coulomb-like'' forces between the branes and the anti-brane, we
find, in addition, an unexpected force between these objects which
also arises from their back-reaction.

The other terms we have computed at second order are the corrections
to the Yang-Mills gauge couplings and theta angles. We find that these
threshold corrections depend on the brane as well as the anti-brane
moduli and lead to a non-holomorphic ``gauge kinetic function''. This
dependence on the anti-brane modulus and the non-holomorphicity has
important consequences for gaugino condensation. This will be the
subject of a forthcoming paper~\cite{otherpapers}.

We believe that some of the features found in this paper, for example
the non-Coulomb type brane-brane forces due to anti-brane back-reaction
and the non-holomorphic corrections to the gauge kinetic function, may
arise in other models with anti-branes, for example, in the context of
IIB string theory. To our knowledge, this has not yet been studied.
Our results have important applications to moduli stabilization in
heterotic models with anti-branes, supersymmetry breaking due to
anti-branes and the cosmology of such models. In particular, given
that the anti-brane can be attracted to the boundary by the
perturbative forces while it is repelled due to membrane instanton
effects, we have argued that stabilization of the anti-brane due to a
combination of perturbative and non-perturbative effects can be
achieved. These issues are currently under
investigation~\cite{otherpapers}.

\vskip 0.5cm
\noindent
{\bf Acknowledgments\\}
A.~L.~is supported by the EC 6th Framework Programme
MRTN-CT-2004-503369.  J.~G.~is supported by CNRS and in part for this
research by the National Science Foundation under Grant
No. PHY99-07949.  The work of B.~A.~O~is supported in part by the DOE
under contract No. DE-AC02-76-ER-03071 and by the NSF Focused Research
Grant DMS0139799.  J.~G.~and A.~L.~would like to thank the Department
of Physics of the University of Pennsylvania, where part of this work
was carried out, for generous hospitality.


\section*{Appendix: Origins of Quantities in the Five-Dimensional
  Theory.}

\appendix

In this appendix we explain the origin of the fields and quantities
which appear in the action~\eqref{5daction} of five-dimensional
heterotic M-theory in terms of the underlying 11-dimensional
theory~\cite{Horava:1996ma}, compactified on a Calabi-Yau manifold
$X$. All of these results are standard and can be found in the
literature~\cite{Candelas:1990pi,Lukas:1998yy,Lukas:1998tt}, but are here included for
completeness. The last part of the appendix presents a number of
technical results for the five-dimensional warping which are needed
for the calculation in the main part of the paper.

\section{Origin of the Five-Dimensional Bulk Theory} \label{App:mod}

The bulk fields in the five-dimensional action~\eqref{5daction}
originate from zero modes of the 11-dimensional metric and three-form
on the Calabi-Yau manifold $X$. Let us start by discussing the metric
moduli.

In addition to the five-dimensional metric, the 11-dimensional metric
gives rise to a number of scalar fields which parametrize the Calabi-Yau
moduli space. As is well known, this space is (locally) a direct product
of the K\"ahler moduli space, parametrized by the periods of the K\"ahler
form $\omega$, and the complex structure moduli space, parametrized by
periods of the holomorphic three-form $\Omega$. Let us first discuss
the K\"ahler moduli space. We introduce an integral basis $C^k$, where
$k,l,\dots =1,\dots ,h^{1,1}(X)$, of the second homology of the Calabi-Yau
manifold and a dual basis $v^{-1/3}\omega_k$ of harmonic $(1,1)$ forms satisfying
\begin{equation}
  \frac{1}{v^{1/3}} \int_{C^k}\omega_l=\delta_l^k\; .
\label{last3}
\end{equation}
The $h^{1,1}(X)$ K\"ahler moduli $a^k$ can then be defined by
expanding $\omega =a^k\omega_k$.  The K\"ahler moduli space metric
$G_{kl}(a)$ is then given by
\begin{equation}
 G_{kl}(a)\equiv\frac{1}{vV}\int_X\omega_k\wedge\star\omega_l
          =\frac{\partial}{\partial a^k}\frac{\partial}{\partial a^l}
            K(a)
\end{equation}
where the K\"ahler potential $K(a)$ is given by $K(a)=- \ln (d_{klm}a^ka^la^m)$ with
the triple intersection numbers
\begin{equation}
 d_{klm}=\frac{1}{v}\int_X\omega_k\wedge\omega_l\wedge\omega_m\; .
\end{equation}
It is easy to show that the total Calabi-Yau volume V can be written as
\begin{equation}
 V=\frac{1}{v}\int_X\sqrt{g}=\frac{1}{6} d_{ijk} a^i a^j a^k\; .
\end{equation}
In the context of the five-dimensional theory it is more appropriate
to work with the rescaled K\"ahler moduli $b^k=V^{-1/3}a^k$ which
satisfy the constraint $e^{-K(b)}=d_{klm}b^kb^lb^m=6$. They constitute
$h^{1,1}(X)-1$ independent fields which will be used in the
five-dimensional action alongside $V$ as the remaining independent
degree of freedom.  Accordingly, the five-dimensional theory is
formulated in terms of the metric $G_{kl}(b)=V^{2/3}G_{kl}(a)$ which
we frequently denote by simply $G_{kl}$ in the text. It is useful to
define lowered index fields $b_k=G_{kl}(b)b^l$. It follows directly
from the explicit form of the metric that $b_kb^k=3$, which, upon
treating the $b^k$ as five-dimensional fields, leads to the identities
\bea
&&b_k \partial_{\alpha} b^k = 0 \; ,\\
&&b_k \left( \nabla_{\alpha} \nabla^{\alpha} b^k +
  \Gamma^k_{lm}(b) \partial_{\alpha} b^l \partial^{\alpha} b^m \right)
=0 \;.  \eea Here, $\Gamma^k_{lm}(b)$ is the connection associated to
the moduli space metric $G_{kl}(b)$ and and $\nabla_\alpha$ is the
five-dimensional covariant derivative. These relations are useful in
deriving the result for the domain wall warping of $b^k$, \eqref{bk}.

\vskip 0.4cm

We now move on to the complex structure moduli space of the
Calabi-Yau.  To this end, we introduce a symplectic basis $(a^A,b_B)$,
where $A,B,\dots =0,\dots , h^{2,1}(X)$ of the third homology and a
dual basis $(\alpha_A,\beta^B)$ of harmonic three-forms satisfying the
standard relations
\begin{equation}
 \int_{X}\a_{B}\w\b^{A}=\int_{a^{A}}\a_{B}=\d_{B}^{A}\; ,\quad
 \int_{X}\b^{A}\w\a_{B}=\int_{b_{B}}\b^{A}=-\d_{B}^{A}
\end{equation}
with all other integrals zero. The complex structure moduli space
can be parametrized by the periods  $({\cal Z}^A, {\cal G}_B)$ defined by
\begin{equation}\label{Operiods}
  {\cal Z}^{A}=\int_{a^{A}}\Omega \; ,\quad
  {\cal G}_{B}=\int_{b_{B}}\Omega \, .
\end{equation}
It turns out, that the periods ${\cal G}_B$ are, in fact, functions
of the ${\cal Z}^A$ and can be obtained as derivatives
${\cal G}_B = \frac{\partial}{{\cal Z}^B} {\cal  G}$ of a holomorphic
pre-potential ${\cal G}$. The complex structure moduli K\"ahler potential
is then given by
\bea
 \cK({\cal Z})=-\mbox{ln}\left(i\int_{X}\O\w\bar{\O}\right)
=-\mbox{ln}\left(\cG_{B}\bar{{\cal Z}}^{B}-\bar{\cG}_{B}{\cal
    Z}^{B}\right)\; .
\eea
The ${\cal Z}^A$ are homogeneous coordinates and the physical degrees
of freedom are identified with the affine coordinates
${\mathfrak z}^a={\cal Z}^a/{\cal Z}^0$, where $a,b,\dots =1,\dots ,h^{2,1}(X)$.
In terms of these physical fields the K\"ahler potential can be written as
\bea
  \cK({\mathfrak z})=-\mbox{ln}\left[2i({\cal G}-\bar{{\cal G}}) -i({\mathfrak
    z}^{a}-\bar{{\mathfrak z}}^{a})\left(\frac{\pt {\cal
        G}}{\pt{\mathfrak z}^{a}}+\frac{\pt \bar{{\cal
          G}}}{\pt\bar{{\mathfrak z}}^{a}}\right)\right]\; ,
\eea
and the associated metric is obtained in the usual way as
${\cal K}_{p \bar{q}} ( {\mathfrak z})=\frac{\partial}{\partial p}
\frac{\partial}{\partial \bar{q}} {\cal K}({\mathfrak z})$. 
The final quantity we need to know about is the matrix $M$ which
appears in some of the five dimensional kinetic terms.
It is defined as follows.
\begin{equation}\label{M-matrix}
  M_{AB}=\bar{\cG}_{AB}+T_{AB}\,,\qquad
  T_{AB}=2i\frac{\mbox{Im}\cG_{AC}{\cal Z}^{C}\;\mbox{Im}\cG_{BD}{\cal Z}^{D}}
  {{\cal Z}^{E}\mbox{Im}\cG_{EF}{\cal Z}^{F}}\; ,
\end{equation}
where ${\cal G}_{AB}$ is the second derivative of the pre-potential ${\cal G}$.

\vskip 0.4cm

Finally, we need to discuss the five-dimensional bulk fields which originate
from the 11-dimensional three-form. The purely external part of this three-form
gives rise to a three-form $C_{\alpha\beta\gamma}$ in the five-dimensional theory
(which can be dualized to a scalar). The internal part, on the other hand,
can be expanded in terms of our symplectic basis $(a^A,b_B)$ and, hence,
gives rise to $2(h^{2,1}(X)+1)$ real scalar fields $\xi^B$ and $\tilde{\xi}_A$.
This completes our discussion of the 11-dimensional origin of the bulk theory
in five dimensions. It is explained in Section~\ref{5d} how these fields are
arranged into five-dimensional super-multiplets.

\section{Origin of the Five-Dimensional Boundary Theories} \label{App:boundaries}

The matter field structure seen in the five dimensional theory comes
from an appropriate dimensional reduction of the ten dimensional $E_8$
super-Yang-Mills theories residing on each of the two orbifold fixed
planes.  Given that we will introduce an arbitrary number, $N$, of M
five-branes in our compactifications, we will label the two boundaries
by $p=0,N+1$.  On both orbifold planes we have stable holomorphic
gauge bundles $V_{p}$ with structure groups $G_{(p)}$ residing on the
Calabi-Yau manifold $X$. The commutants $H_{(p)}$ of $G_{(p)}$ within
$E_8$ are the low-energy gauge groups which appear on the
four-dimensional boundaries of five-dimensional heterotic M-theory. In
order to discuss matter fields we need to decompose the adjoint
representation of $E_8$, ${\bf 248}_{E_8}$, under $G_{(p)}\times
H_{(p)}$.  We write
\begin{equation}
 {\bf 248}_{E_8}=\sum (S_{(p)},R_{(p)})\; , \label{248}
\end{equation}
where $S_{(p)}$ and $R_{(p)}$ denote representations of $G_{(p)}$ and
$H_{(p)}$, respectively. On the four-dimensional boundaries we have,
in addition to gauge fields with gauge group $H_{(p)}$, matter fields
transforming in the representations $R_{(p)}$ which appear in the
decomposition~\eqref{248}. Their number is given by
$h^1(X,V_{R_{(p)}})$, where $V_{R_{(p)}}$ is the associated bundle in
the representation $R_{(p)}$. We will denote these matter fields by
$C_{(p)}^{Ix}$ where indices $I,J,\dots$ run over all irreducible
$R_{(p)}$ representations and indices $x,y,\dots $ label states within
each representation. Further, we denote by $u_I^x$ a basis of
one-forms for the cohomology groups $H^1(X,V_{R_{(p)}})$. The matter
field K\"ahler metric on the four-dimensional boundaries can then be
formally defined as
\begin{equation}
    G_{(p)IJ}=\frac{1}{vV^{2/3}}\int_{X}u_{I}^{x}\w\star\bar{u}_{J x} \;.
\end{equation}
The connection $\Gamma_{(p)IJ}^k$ which appears in the five-dimensional Bianchi
identities can then be written as
\begin{equation}
\Gamma^{k}_{(p)IJ} = G^{kl}(b) \frac{\partial}{\partial b^l} G_{(p)IJ} \;.
\end{equation}
For the Yukawa couplings we can write
\begin{equation}
\lambda_{IJK}=\frac{1}{\|\Omega \|^{2}}\int_{X}\Omega \wedge
u_{I}^{x}\wedge u_{J}^{y}\wedge u_{K}^{z} f_{xyz}^{(IJK)}\; ,
\end{equation}
where $f_{xyz}^{(IJK)}$ projects onto the singlet of the internal gauge group 
$G_{(p)}$. The superpotentials $W_{(p)}$ are then given by
\begin{equation}
 W_{(p)}=\sqrt{4\pi\alpha_{\rm GUT}}\sum_{I,J,K}\lambda_{IJK}f_{xyz}^{(IJK)}
        C_{(p)}^{Ix}C_{(p)}^{Jy}C_{(p)}^{Kz} \;. \label{W}
\end{equation}

\section{Origin of Three-Brane World-Volume Theories} \label{App:5brane}

As explained in the main part of the paper, we are interested in vacua
with $N$ M five-branes, of which $N-1$ are ordinary five-branes and
the remaining one is an anti five-brane. We label these branes by
indices $p,q,\dots = 1,\dots ,N$ with $p=\bar{p}$ corresponding to the
anti five-brane, as indicated in Fig.~\ref{fig1}. The five-branes,
$p\neq\bar{p}$, are wrapped on holomorphic curves with effective
homology classes $C^{(p)}=\beta_k^{(p)}C^k$ within the Calabi-Yau
manifold $X$, where $(C^k)$ is the integral basis of the second
homology of $X$ introduced earlier. The anti five-brane can be viewed
as a five-brane wrapping a holomorphic curve with the ``wrong''
orientation and, hence, its class
$C^{(\bar{p})}=\beta_k^{(\bar{p})}C^k$ corresponds to an
anti-effective class (that is, $-C^{(\bar{p})}$ is effective).

In five-dimensional heterotic M-theory these five-branes appear as
three-branes and we should discuss the fields on their
world-volumes. From the 11-dimensional embedding coordinates of the
five-branes we first have the position moduli $y_{(p)}$ in the
orbifold direction and, possibly, additional scalar fields which
describe the moduli space of the curves $C^{(p)}\subset X$. In this
paper, we will not consider the latter fields explicitly. The purely
internal part of the two-form on the five-brane world-volume (both
indices in the direction of the curves) give rise to axions $s_{(p)}$
on the three-branes\footnote{Due to the self-duality of the five-brane
  two-form its purely external part does not contribute any new
  degrees of freedom.}. Further let us introduce a standard basis
$(a_{(p)u},b_{(p)w})$ of A and B cycles, where $u,w,\dots = 1,\dots
,g_{(p)}$ and $g_{(p)}$ is the genus of the $p^{\rm th}$ curve.  In
addition, we introduce a basis of holomorphic one-forms $(\a_{(p)u})$
on each curve, satisfying
\begin{equation}
\int_{a_{(p)u}}\a_{(p)w}=\delta_{uw}\; .
\end{equation}
We can then expand the five-brane two form
in this basis of one-forms which gives rise to $2g_{(p)}$ Abelian vector fields
on the three-brane world-volumes. However, due to the self-duality of the
five-brane two-form only half of these vector fields are independent and we
will denote their field strengths by $E^u_{(p)\mu\nu}$. The period matrices
\begin{equation}
 \Pi_{(p)uw}\equiv\int_{b_{(p)u}}\a_{(p)w}
\end{equation}
then determine the gauge kinetic functions of these vector fields.

\section{Solution of the Bianchi Identities} \label{App:Bianchi}

In this appendix we collect the results for the warping of the
five-dimensional bulk fields due to field fluctuations on the
boundaries.  These equations, which correspond to the 11-dimensional
warping results obtained in Ref.~\cite{Lukas:1997fg}, are needed for
the reduction from five to four dimensions in the main part of the
paper.

We start by solving the Bianchi identities~\eqref{BI1}--\eqref{BI3}.
The Bianchi identity for the ${\cal F}^k$'s, \eqref{BI2}, can be
solved yielding the following result.  \bea {\cal F}^k_{\mu \nu} &=&
(-2 \kappa^2_5) J^{(0) k}_{2\mu \nu} + 2 \kappa^2_5 z( J^{(0) k}_{2\mu
  \nu} + J^{(N+1) k}_{2\mu \nu}) \\ {\cal F}^k_{\mu y} &=& \frac{-2
  \kappa^2_5}{\pi \rho} (\omega^{(0) k}_{1 \nu} +\omega^{(N+1)
  k}_{1\nu} ) + \tilde{{\cal F}}^k_{\mu y} \eea The tilded quantity,
$\tilde{{\cal F}}$, is an unspecified piece with trivial Bianchi
identity, $d \tilde{{\cal F}} =0 $, which can be fixed upon an
examination of the equations of motion. We have also introduced the
Chern-Simons one forms $\omega^{(p)k}$ which are defined by the
relation $d \omega^{(p) k}_1 = J^{(p) k }_2$, where $J^{(p) k}_2$ was
defined in \eqref{mattercurrent}.  Part of $\tilde{{\cal F}}$
is that due to the zero mode $\chi^k = {\cal A}^k_y$. We also note
that ${\cal A}_{\mu}$, being odd under the ${\mathbb Z}_2$ orbifold
action, does not lead to zero modes. It turns out that the rest of the
warping of ${\cal F}$ is not necessary for our present purposes. It
drops out from the calculation of the four dimensional effective
action.

\vspace{0.1cm}

The next set of terms from the five dimensional action that we need to
dimensionally reduce are those involving the four-form field strength $G$. 
The solution to its Bianchi identity \eqref{BI1} is given by
\bea
\label{bis1} G_{\mu \nu \rho \gamma} &=& -2 \kappa^2_5 J^{(0)}_{4\mu
  \nu \rho \gamma} + 2 \kappa^2_5 z( J^{(0)}_{4\mu \nu \rho \gamma} +
J^{(N+1)}_{4\mu \nu \rho \gamma}) + \tilde{G}_{\mu \nu \rho
  \gamma}\\ \label{bis2} G_{\nu \rho \gamma y} &=& \frac{-2
  \kappa^2_5}{\pi \rho} \frac{1}{16 \pi \alpha_{\textnormal{ GUT}}}
(\omega^{(0)}_{3 \nu \rho \gamma}+ \omega^{(N+2)}_{3 \nu \rho \gamma})
+ \tilde{G}_{\nu \rho \gamma y} \eea Here $\tilde{G}$ is an
unspecified contribution to the four form field strength which must be
fixed upon an examination of the equations of motion. It obeys the
trivial Bianchi Identity: $d \tilde{G} =0$. The quantities
$\omega_{3}^{(p)}$ in \eqref{bis1} are Chern-Simons three-forms
defined by the relations $d \omega^{(p)}_{3} = F_{(p)} \wedge
F_{(p)}$.  As a check of the integrability of this result we can find
the associated three form potential contributions.
\bea \label{Cw3}C_{\mu \nu \rho} = \frac{-2 \kappa^2_5}{16 \pi
  \alpha_{\textnormal{GUT}}} ( \omega^{(0)}_{3 \mu \nu \rho} - z
(\omega^{(0)}_{3 \mu \nu \rho}+\omega^{(N+1)}_{3 \mu \nu \rho})) \eea
The zero mode $B_{\mu\nu}=C_{y\mu\nu}$ with field strength
$H_{\mu\nu\rho}$ is contained in the $\tilde{G}$ part of the above
solution, along with other contributions to the warping which are
irrelevant for our calculation.

Finally, we need to consider the Bianchi identity~\eqref{BI3}
for ${\cal X}^A$ and $\tilde{\cal X}_B$. In the absence of flux
the solution to these Bianchi identities only give rise to higher
order terms in the four-dimensional effective theory and are not
explicitly needed.

\vskip 0.4cm

For our calculation of the gauge kinetic functions to second order in
subsection~\eqref{threshold} we need explicit expressions for the
warping of the metric and the Calabi-Yau volume modulus $V$ caused by
field fluctuations on the boundaries. These warpings from fluctuations
have to be added to the domain wall solution in section~\ref{dw} to
obtain the full expressions needed in the calculation.  \bea
 \label{metricf2warping}
\delta g_{\mu \nu} &=&\frac{\rho\kappa_5^2}{2\a_{\rm GUT}}\frac{b_0V_0}{a_0^2}
   \left[\left(\frac{1}{2}z^2-z+\frac{1}{3}\right){\rm tr}(F_{(0)\mu\gamma}
   {F_{(0)\nu}}^\gamma )+\frac{1}{2}\left(z^2-\frac{1}{3}\right){\rm tr}
  (F_{(N+1)\mu\gamma} {F_{(N+1)\nu}}^\gamma )\right]\\
\delta a &=& -\frac{\rho\kappa_5^2}{16\a_{\rm GUT}}\frac{b_0V_0}{a_0^3}
             \left[\left(\frac{1}{2}z^2-z+\frac{1}{3}\right){\rm tr}(F_{(0)}^2)
             +\frac{1}{2}\left( z^2-\frac{1}{3}\right){\rm tr}(F_{(N+1)}^2)\right]\\
\delta V &=& \frac{2V_0}{a_0}\delta a
\eea


\end{document}